\documentclass[aps,pre,superscriptaddress,noshowpacs]{revtex4}
        \usepackage{graphicx}
        \usepackage{dcolumn}
        \usepackage{bm} 
        \usepackage{epstopdf}
        \usepackage{color}
        \usepackage{amsmath}
        \usepackage{amssymb}
        
        \newcommand{\refeq}[1]{\mbox{Eq.~(\ref{#1})}}

        \newcommand{\be}{\begin{equation}}
        \newcommand{\ee}{\end{equation}}
        \newcommand{\ba}{\begin{eqnarray}}
        \newcommand{\ea}{\end{eqnarray}}

\def\II{\hbox{$1\hskip -1.2pt\vrule depth 0pt height 1.6ex width 0.7pt\vrule depth 0pt height 0.3pt width 0.12em$}}
\begin{document}
\title{Experimental investigation of the fluctuations in nonchaotic scattering in microwave billiards}
\author{Runzu Zhang}
        \affiliation{School of Physical Science and Technology, and Key
        Laboratory for Magnetism and Magnetic Materials of MOE, Lanzhou University, Lanzhou, Gansu 730000, China}
\author{Weihua Zhang}
        \affiliation{School of Physical Science and Technology, and Key
        Laboratory for Magnetism and Magnetic Materials of MOE, Lanzhou University, Lanzhou, Gansu 730000, China}
\author{Barbara Dietz}
        \email{dietz@lzu.edu.cn}
        \affiliation{School of Physical Science and Technology, and Key
        Laboratory for Magnetism and Magnetic Materials of MOE, Lanzhou University, Lanzhou, Gansu 730000, China}
\author{Chai Guozhi}
        \affiliation{Key
        Laboratory for Magnetism and Magnetic Materials of MOE, Lanzhou University, Lanzhou, Gansu 730000, China}
\author{Liang Huang}
        \affiliation{School of Physical Science and Technology, and Key
        Laboratory for Magnetism and Magnetic Materials of MOE, Lanzhou University, Lanzhou, Gansu 730000, China}
\date{\today}
\begin{abstract}
We report on the experimental investigation of the properties of the eigenvalues and wavefunctions and the fluctuation properties of the scattering matrix of closed and open billiards, respectively, of which the classical dynamics undergoes a transition from integrable via almost integrable to fully chaotic. To realize such a system we chose a billiard with a $60^\circ$ sector shape of which the classical dynamics is integrable, and introduced circular scatterers of varying number, size and position. The spectral properties of generic quantum systems of which the classical counterpart is either integrable or chaotic are universal and well understood. If, however, the classical dynamics is pseudo-integrable or almost-integrable, they exhibit a non-universal intermediate statistics, for which analytical results are known only in a few cases, like, e.g., if it corresponds to semi-Poisson statistics. Since the latter is, above all, clearly distinguishable from those of integrable and chaotic systems our aim was to design a billiard with these features which indeed is achievable by adding just one scatterer of appropriate size and position to the sector billiard. We demonstrate that, while the spectral properties of almost-integrable billiards are sensitive to the classical dynamics, this is not the case for the distribution of the wavefunction components, which was analysed in terms of the strength distribution, and the fluctuation properties of the scattering matrix which coincide with those of typical, fully chaotic systems.
\end{abstract}
\maketitle

\section{\label{sec:level1}Introduction}
The main focus of the field of quantum chaos is the search for signatures of classical chaos in properties of the eigenvalues and wavefunctions of the corresponding bounded quantum system or those of the scattering matrix in open ones. While the characteristics are well understood by now for the former, many questions remain open for scattering systems. Planar billiards -- bounded, simply connected domains in which a pointlike particle moves freely and is reflected back specularly on impact with the boundary -- provide a suitable system for the investigation of manifestations of classical chaos in the corresponding quantum system, since their classical dynamics only depends on their shape~\cite{Sinai1970,Bunimovich1991,Berry1981}. Examples for integrable systems are rectangular, circular and elliptic billiards, whereas the classical dynamics of the Bunimovich stadium~\cite{Bunimovich1979} or the Sinai billiard~\cite{Sinai1970} is chaotic except for a set of measure zero in phase space. 
	
It is known by now and has been confirmed numerically and also experimentally that the spectral properties of generic integrable and fully chaotic systems are universal. Those of the former are well described by those of random numbers generated in a Poisson process~\cite{Berry1977a} and coincide for the latter with those of the eigenvalues of random matrices from the Gaussian orthogonal ensemble (GOE) for generic quantum systems with a fully chaotic classical dynamics and preserved time-reversal invariance~\cite{Berry1977,Casati1980,Bohigas1984,Mehta1990}. Furthermore, the spectral properties of systems with mixed regular-chaotic dynamics may posses generic features and accordingly be described by random matrix ensembles interpolating between the two extreme cases~\cite{Rosenzweig1960,Berry1984,Lenz1991,Kota2014}. 
	
Billiards with the shapes of rational polygonals containing corners with angles $\alpha\ne\frac{\pi}{n}$ where $n$ is an integer~\cite{Richens1981,Zyczkowski1992,Zyczkowski1994,Biswas1990,Shudo1993,Shudo1994} and of an integrable one containing pointlike scatterers~\cite{Seba1990,Haake1991,Seba1991,Shigehara1993,Shigehara1994,Shigehara1996,Cheon1996,Weaver1995,Legrand1997,Rahav2002,Rahav2003} or, in general, an obstacle of a size which is much smaller than the billiard area and smaller or comparable to the wavelength of the quantum particle trapped in it, are pseudo-integrable and almost-integrable, respectively. Pseudo-integrable and almost-integrable systems exhibit an intermediate spectral statistics~\cite{Bogomolny1999,Bogomolny2001} which generally is non-universal. In Refs.~\cite{Stoeckmann1990,Exner1997,Bogomolny2001a,Bogomolny2002,Tudorovskiy2008,Tudorovskiy2010,Tudorovskiy2011,Bialous2016} the spectral properties of billiards with the shape of an integrable one containing a $\delta$-function potential, called singular billiards, were investigated theoretically and experimentally, revealing that they exhibit a non-universal singular statistics. 
	
These drastical changes of the spectral properties, caused by introducing a singular scatterer or a $\delta$-function potential, were at first sight surprising because their classical dynamics is integrable or almost-integrable in the sense that the trajectories which hit the singular vertices or the pointlike scatterer are of measure zero or negligibly small in phase space, respectively, and have been investigated intensively theoretically and also experimentally. However, the properties of scattering systems, of which the internal dynamics is nonchaotic have been scarcely investigated~\cite{doCarmo2019}. The focus of the present article is their experimental investigation using a microwave billiard. 
	
In Sec.~2 we will outline how we realized a quantum billiard with integrable, almost-integrable and chaotic dynamics and in Sec.~3 the experimental setup will be introduced. In Sec.~4 the experimental results for the spectral properties will be presented and, finally, in Sec.~5 those for the fluctuation properties of the associated scattering matrix will be discussed.    

\section{\label{QB} Quantum billiard}
	In order to achieve a quantum system with integrable, almost-integrable and chaotic dynamics we use a billiard with the shape of a $60^\circ$ circle sector and introduce circularly-shaped scatterers of increasing size and number. The classical dynamics of a circle-sector shaped billiard~\cite{Dietz1995} is integrable, the constants of motion being energy and angular momentum. The Schr\"odinger equation of the corresponding quantum billiard is given by the Laplacian in polar coordinates $(r,\varphi)$ inside the bounded domain $\Omega$ with Dirichlet conditions along the boundary $\partial\Omega$, 
\ba
\left[\Delta_{r,\varphi}+k^2\right]\psi\left(\boldsymbol{r}\in\Omega\right)=0,\,\psi(\boldsymbol{r}\in\partial\Omega)=0,\\
\Omega=\left\{(r,\varphi):\, 0\leq r < R,\, 0 < \varphi < \theta\right\}.\nonumber
\label{Schr}
\ea
The wavefunctions are given in terms of Bessel functions, 
\be
\psi_{m,\nu}(r,\varphi)=\sin\left(\frac{m\pi}{\theta}\varphi\right)J_{\frac{m\pi}{\theta}}(k_{m,\nu} r),
\label{wavefunctions}
\ee
with $k$ denoting the wavevector. They were defined such that they vanish along the straight part of the sector-shaped boundary. The eigenwavevectors $k_{m,\nu}$ are determined by imposing the Dirichlet boundary condition on the curved part, $\psi_{m,\nu}(r=R,\varphi)$, yielding the eigenvalue equation
\be
J_{\frac{m\pi}{\theta}}(k_{m,\nu}R)=0.
\label{eigenvalues}
\ee
We turn the classical dynamics of the sector billiard into an almost-integrable one by inserting a pointlike circular scatterer with Dirichlet boundary conditions along the walls, which corresponds to a pointlike hole, at a suitable position. Our aim was to achieve a quantum billiard with the spectral properties exhibiting intermediate statistics. To be more explicit, we seeked a quantum billiard of which the spectral properties exhibit intermediate statistics close to semi-Poisson statistics~\cite{Bogomolny1999,Rahav2003}, and thus were well distinguishable from Poisson and GOE statistics. Furthermore, for this case analytical expressions are available for the relevant statistical measures. A quantum system of which the spectral properties are close to GOE statistics was realized by adding three circular scatterers to the sector billiard. We performed numerical simulations to find appropriate sizes and positions of the scatterers and accordingly selected those listed in Tabs.~\ref{tab:one}~I and~\ref{tab:three}~II, respectively. The positions of the scatterers were chosen in the vicinity of the curved boundary where the wavefunctions are generally non-vanishing, and thus will be affected by the scatterer. 
\begin{table*}
\begin{center}
\caption{
Sector billiard with one scatterer. Radius (1st row) and position (2nd row) of the circular scatterer. It was inserted into the $60^\circ$-sector billiard in order to achieve an almost-integrable dynamics, such that the spectral properties of the corresponding quantum billiard were close to semi-Poisson statistics. The radius of the sector equaled $R=800$~mm.}
\vspace{5mm}
\begin{tabular}
      {@{\extracolsep{00pt}}
        @{\extracolsep{10pt} }{l}
        @{\extracolsep{10pt} }{c}
        @{\extracolsep{10pt} }{c}
        @{\extracolsep{10pt} }{c}
        @{\extracolsep{10pt} }{c}
      }%
\hline\hline
\noalign{\smallskip}
Billiard &1st & 2nd & 3rd & 4th\\
\noalign{\smallskip}
\hline
Radius&$0.025R $ & $0.03R$ & $0.03R$ & $0.025R$\\
Position (x,y) [mm]&$\left (640,80  \right )$&$ \left (640,80  \right )$& $\left (640,400  \right ) $& $\left (640,400  \right ) $\\
\hline\hline
\end{tabular}
\label{tab:one}
\end{center}
\end{table*}
\begin{table*}
\begin{center}
\caption{
Sector billiard with three scatterers. Radii and positions of the circular scatterers. They were inserted  into the $60^\circ$-sector billiard in order to achieve a quantum billiard of which the spectral properties follow GOE statistics.
}
\vspace{5mm}
\begin{tabular}
      {@{\extracolsep{00pt}}
        @{\extracolsep{10pt} }{l}
        @{\extracolsep{10pt} }{c}
        @{\extracolsep{10pt} }{c}
        @{\extracolsep{10pt} }{c}
        @{\extracolsep{10pt} }{c}
      }%
\hline\hline
\noalign{\smallskip}
Billiard &Position $(x,y)$ [mm]&$\left (640,400  \right )$&$ \left (520,520  \right )$& $\left (640,80  \right ) $\\
\noalign{\smallskip}
1st &Radii&$0.02R$ & $0.03R$ & $0.05R$ \\
2nd &Radii&$0.02R$ & $0.04R$ & $0.05R$ \\
3rd &Radii&$0.01R$ & $0.04R$ & $0.05R$ \\
\noalign{\smallskip}
	&Position $(x,y)$ [mm]&$\left (640,400  \right )$& $\left (520,520  \right )$& $\left (520,80  \right ) $\\
\noalign{\smallskip}
4th&Radii&$0.02R$ & $0.04R$ & $0.05R$ \\
5th&Radii&$0.025R$ & $0.04R$ & $0.05R$ \\
\noalign{\smallskip}
	&Position $(x,y)$ [mm]&$\left (680,200  \right )$& $\left (520,520  \right )$& $\left (520,80  \right )$ \\
\noalign{\smallskip}
6th&Radii&$0.025R$ & $0.04R$ & $0.05R$\\
\hline\hline
\end{tabular}
\label{tab:three}
\end{center}
\end{table*} 

To attain a scattering system, the quantum billiard is coupled to its environment via single-mode scattering channels~\cite{Albeverio1996,Haake1996,Stoeckmann1998}. Before studying the fluctuation properties of the scattering matrix describing the associated scattering process, we analyzed the spectral properties of the empty sector billiard and those containing one or three circular scatterers. Here, the positions and radii of the scatterers were chosen such that the quantum billiard was almost-integrable and exhibited intermediate statistics close to semi-Poisson statistics~\cite{Bogomolny1999}, or such that the spectral properties were close to those typical for chaotic systems, i.e., followed GOE statistics, respectively. 

For the analysis of spectral properties and the comparison with random matrix theory (RMT) predictions applicable to generic systems exhibiting universal properties, system specific properties need to be extracted, that is, the eigenwavevectors need to be unfolded such that the spectral density is uniform and thus the mean spacing is constant. In quantum billiards this is achieved with Weyl's law for the integrated spectral density, i.e., the number of eigenwavevectors below a given value $k$, 
\be
N^{\rm Weyl}(k)=\frac{A}{4\pi}k^2+\frac{\mathcal{L}}{4\pi}k+C,
\label{Weyl}
\ee
where $A$ and $\mathcal{L}$ denote the area and the perimeter of the billiard, respectively. The eigenwavevectors were unfolded to unity by means of $\epsilon_n=N(k_n)$ with $k_1\leq k_2\leq\cdots$ denoting the sorted-by-size eigenwavevectors of~\refeq{Schr}. For the characterization of the spectral properties we considered the distribution $P(s)$ of nearest-neighbor spacings $s_n=\epsilon_{n+1}-\epsilon_n$ and the cumulative nearest-neighbor spacing distribution $I(s)$ in order to attain information on short range correlations. For random matrices from the GOE $P(s)$ is very well approximated by the Wigner distribution
\be
P^{\rm Wigner}(s)=\frac{\pi}{2}se^{-\frac{\pi}{4}s^2},
\label{Wigner}
\ee
whereas for Poissonian random numbers it is given by
\be
P^{\rm Poisson}(s)=e^{-s}.
\label{Poisson}
\ee
These distributions reflect an inherent difference between generic integrable and fully chaotic systems, namely, the probability that the spacing between two eigenvalues is zero or much less than the mean spacing is maximal for the former ones and vanishingly small for the latter ones. For large spacings, on the other hand, $P(s)$ decays exponentially and Gaussian-like, respectively. These features, namely the linear increase for $s\simeq 0$ and the exponential decay for $s\to\infty$ are combined in the case of semi-Poisson statistics, where  the nearest-neighbor spacing distribution is given by
\be
P(s)=4se^{-2s}.
\ee
We also analyzed long-range correlations in terms of the number variance $\Sigma^2(L)=\langle (N(L)-L)^2\rangle$ in an interval of length $L$, where due to the unfolding $\langle N(L)\rangle =L$, and the Dyson-Mehta statistics,
\be
\Delta_3(L)=\left\langle\underset{a,b}{\rm min}\int_{E-L/2}^{E+L/2}\left[ N(E)-a-bE\right]^2{\rm d}E\right\rangle, 
\ee
which provides a measure for the spectral rigidity. 

\section{Experimental setup\label{MB}}
\begin{figure*}
\begin{center}
\includegraphics[width=7.6cm,height=4.6cm]{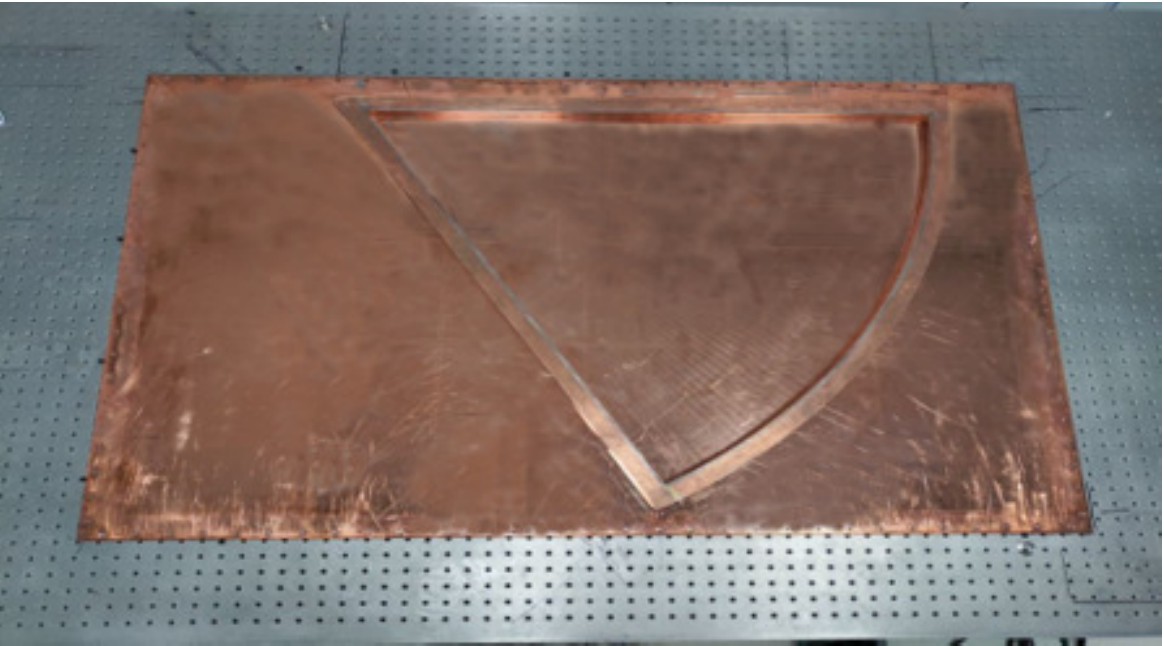}
\includegraphics[width=7.6cm,height=4.6cm]{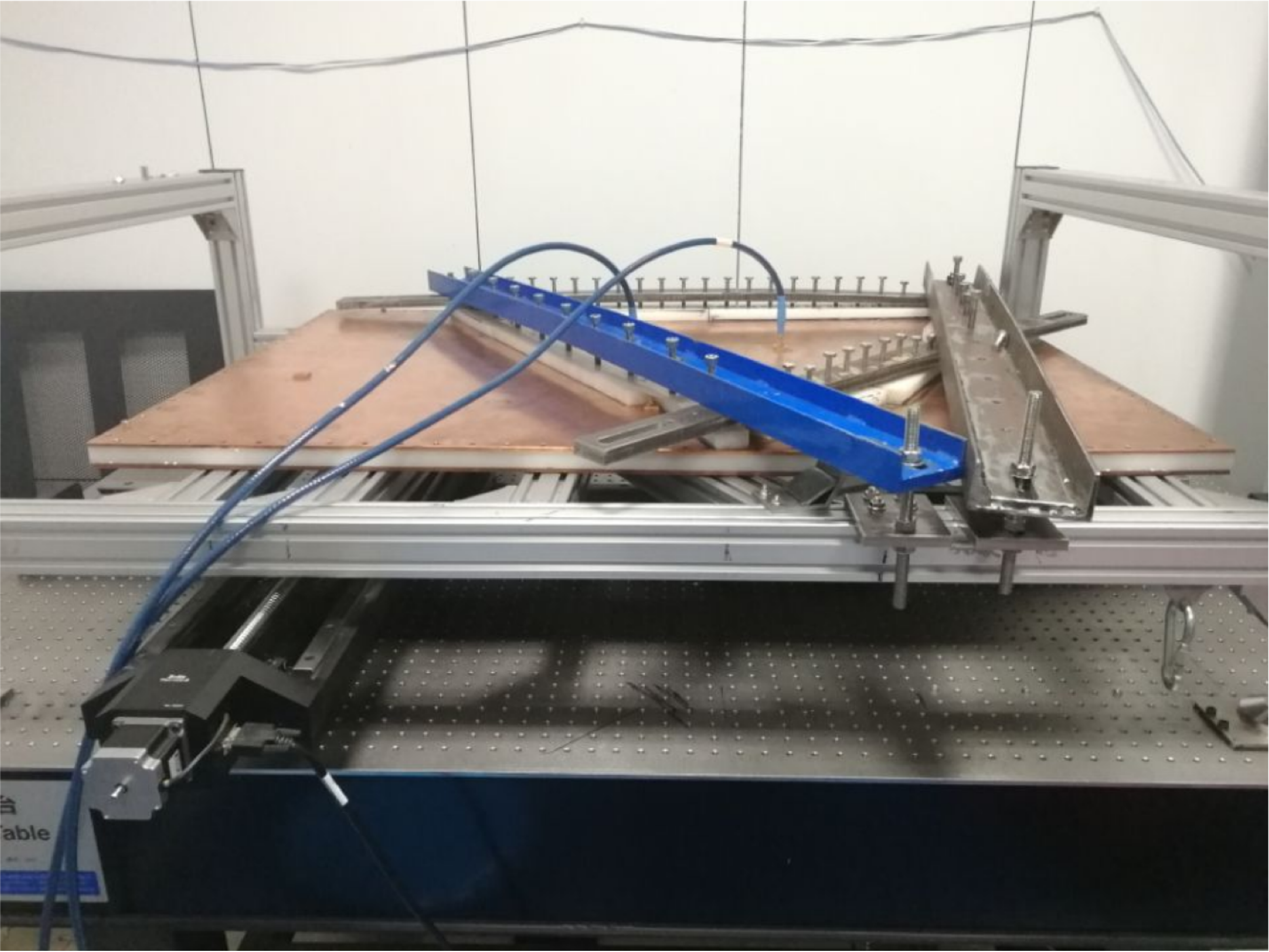}
\caption{
(color online) Photograph of the bottom plate and the frame defining the walls of the microwave billiard (top) and of the resonator (bottom) which is composed of the frame squeezed between the bottom and top plate as sketched in Fig.~\ref{fig2}. All three parts are made from copper. The size of the plates is 1260$\times$860$\times 5$~mm. The frame has the shape of a $60^\circ$ circle sector of radius 800~mm and height 20~mm corresponding to a cutoff frequency $f_{\rm max}\simeq 7.5$~GHz. In order to ensure a good electrical contact, all parts were squeezed together tightly with screw clamps as illustrated in the bottom figure. Furthermore, a rectangular frame of the same size as the plates and height as the frame, and the top and bottom plates were firmly screwed together along their edges through screw holes in order to improve the electrical contact. For the measurement of the resonance spectra two antennas were attached to the resonator and coupled to the VNA via cables visible in the right figure. Furthermore, the resonator rested on a frame and a positioning unit recognizable in the left bottom part guided the perturbation body with a magnet which was attached to it from underneath the resonator along the bottom plate.}\label{fig1}
\end{center}	
\end{figure*}

\begin{figure*}
\begin{center}
\includegraphics[width=7.6cm,height=4.6cm]{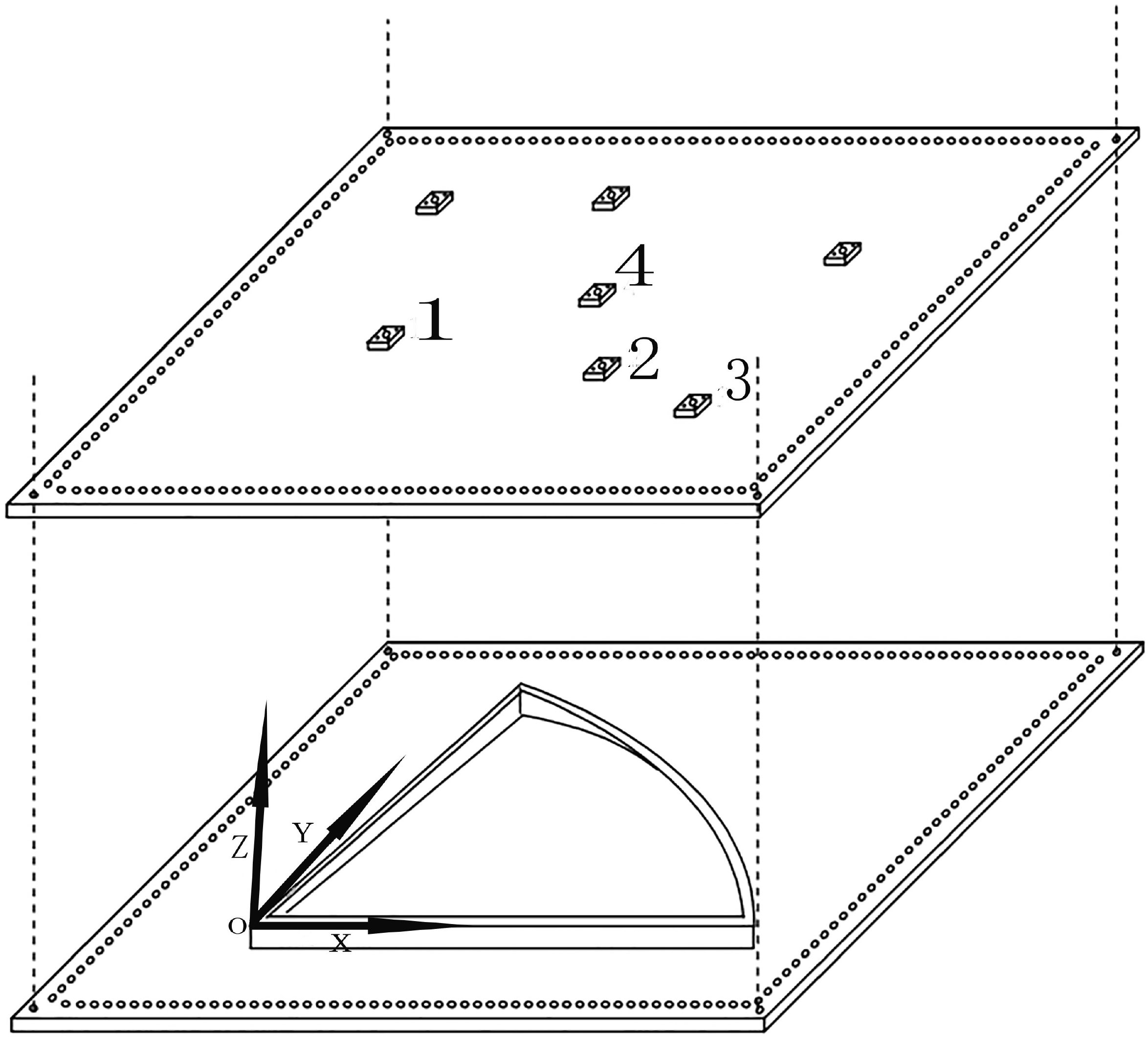}
\caption{
Schematic view of the three parts forming the microwave billiards. A rectangular frame of the same size as the top and bottom plate was added and all three parts were screwed together through holes in order to improve the electrical contact. The sector frame had grooves along the inner rim of its top and bottom surface into which wire of solder was placed to ensure this in the whole frequency range~\cite{Bogomolny2006}. Seven antenna ports are fixed to the top plate of which four, marked by numbers, were used in the experiments. 
}\label{fig2}
\end{center}
\end{figure*}

The experiments were performed with a flat, metallic microwave resonator with the shape of a $60^\circ$ sector. Here we employ the equivalence of the Helmholtz equation governing it and the non-relativistic Schr\"odinger equation of the quantum billiard of corresponding shape. This analogy holds for frequencies $f$ of the microwaves inside the cavity below a cutoff frequency $f\leq f_{\rm max}=c_0/2h$ with $c_0$ denoting the velocity of light and $h$ the height of the resonator, where the electric field vector is perpendicular to the top and bottom plate, so that the Helmholtz equation for $\vec E(\vec r)=E_{\rm z}(x,y)\vec e_{\rm z}$ becomes scalar and is mathematically identical with the Schr\"odinger equation~\refeq{Schr}. Accordingly, we refer to such resonators as microwave billiards~\cite{StoeckmannBuch2000,Richter1999}. Due to this analogy the eigenwavevectors $k_n$ and wavefunctions $\psi_n(\vec r)$ of a quantum billiard can be determined experimentally from the eigenfrequencies $f_n =\frac{c_0k_n}{2\pi}$ and the electric field distribution $E_{\rm z,n}(x,y)$ at frequency $f=f_n$. Figures~\ref{fig1} and~\ref{fig2} show a photograph and a schematic view of the microwave billiard, which had the shape of a $60^\circ$ circle sector. It was composed of three parts, a top and bottom plate and a frame defining the shape of the resonator which were made from copper. Note, that in distinction to previous experiments~\cite{Bogomolny2006,Dietz2007}, where the top and bottom plates where screwed together through holes along the frame of the resonator in order to achieve a good electrical contact, our construction avoids the screw holes and uses screw clamps instead so that other billiard shapes may be realized by simply replacing the frame.  

The eigenfrequencies correspond to the positions of the resonances in the reflection and transmission spectra of the microwave billiard.  They were measured by attaching two antennas at two of four possible ports located at the positions marked in Fig.~\ref{fig2} which were connected to an Agilent N5227A vector network analyzer (VNA) via SUCOFLEX126EA/11PC35/1PC35 coaxial-cables emitting microwave power into the resonator via one antenna and receiving it at the same or the other one, respectively. The VNA yielded as output the relative phases $\phi_{\rm ba}$ and the ratios of the microwave power, $\frac{P_{\rm out, b}}{P_{\rm in, a}}=\vert S_{\rm ba}\vert^2$ of the rf signal sent into the resonator at antenna $a$ and coupled out at antenna $b$ and thus the complex scattering matrix element $S_{\rm ba}=\vert S_{\rm ba}\vert e^{i\phi_{\rm ba}}$ describing this scattering process~\cite{Dietz2008,Dietz2009,Dietz2010}. In the vicinity of an isolated or weakly overlapping resonance at $f=f_n$ $S_{\rm ba}$ is well described by the complex Breit-Wigner form   
\begin{equation}
S_{\rm ba}=\delta_{\rm ba}-i \frac{\sqrt{\Gamma_{n a} \Gamma_{n b}}}{f-f_{n}+\frac{i}{2} \Gamma_{n}}.
	\label{Breit}
\end{equation}
Here, $\Gamma_{\rm n a}$ and $\Gamma_{\rm n b}$ are the partial widths associated with antennas $a$ and $b$, and $\Gamma_{\rm n}$ is the total width of the resonance which is given by the sum of the partial widths of the emitting and receiving antennas and the width $\Gamma_{\rm abs}$ due to the absorption in the walls of the resonator, $\Gamma_n=\Gamma_{\rm n a}+\Gamma_{\rm n b}+\Gamma_{\rm abs}$ which, actually, provides the dominant contribution in measurements at room temperature. We would like to emphasize that, only if the resonances are at most weakly overlapping, that is, as long as~\refeq{Breit} is applicable to the resonance spectra, the eigenfrequencies of the microwave billiard coincide with the eigenvalues of the closed quantum billiard of corresponding shape.

\begin{figure*}
\begin{center}	
\includegraphics[width=7.2cm,height=5.1cm]{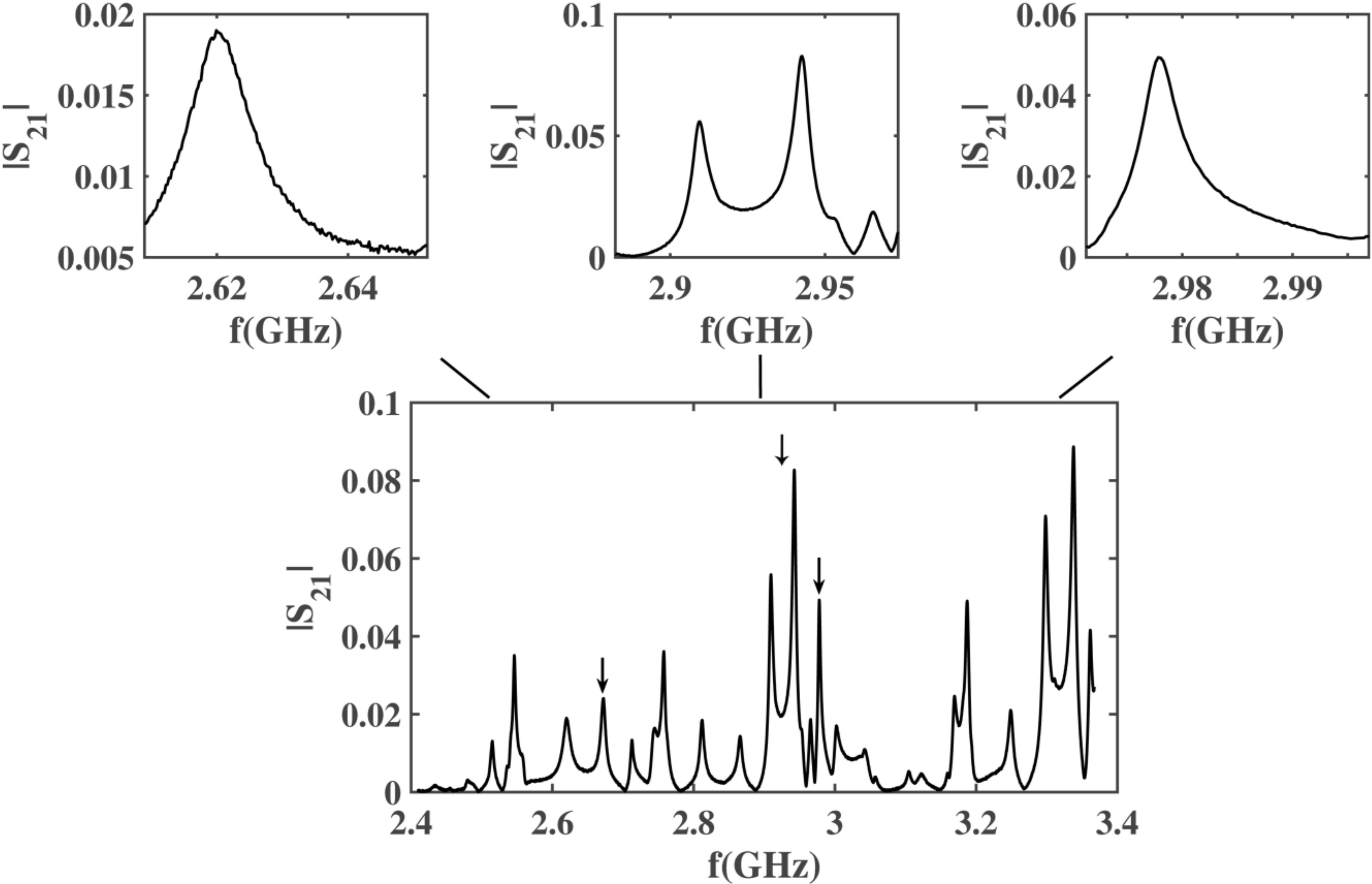}
\caption{
Part of a transmission spectrum of the empty microwave sector billiard. The positions of the resonances yield the eigenfrequencies of the microwave billiard, that is, the eigenvalues of the corresponding quantum billiard. The zooms into the spectrum show two isolated resonances (left and right) and a pair of weakly overlapping ones (middle). The measurement of the electric field intensity distribution at either of the two eigenfrequencies of the latter yields a superposition of both distributions. Therefore, we only measured distributions at isolated resonances like those depicted in this figure.} 
\label{fig:spec}
\end{center}
\end{figure*}

The eigenfrequencies, and thus the eigenvalues of the corresponding quantum billiard, are determined by fitting the complex Breit-Wigner form~\refeq{Breit} to the measured scattering matrix elements. For this to be feasible it is crucial that the widths of the resonances are small compared to the average spacing between adjacent resonances. Consequently, a cavity with a high-quality factor $Q$ of the resonator is a prerequisite. The $Q$ factor depends on the material of the resonator, or to be more explicit, on the size of the absorption of microwave power in its walls, which leads to a broadening of the widths of the resonances and thus to an overlapping of neighboring ones. High $Q$ factors of up to $Q\simeq 10^7$ were achieved in measurements at liquid-helium temperature $T_{\rm LHe}=4$~K with microwave billiards made from niobium or coated with lead, which are superconducting at this temperature~\cite{Dietz2015a}. Our experiments, however, were done at room temperature and thus we had to cope with overlapping resonances. To reduce absorption, we constructed the cavity from high quality copper. The $Q$ value, in addition, is proportional to the ratio of the volume to the surface of the resonator, that is, essentially to its height $h$. Yet, the analogy between the quantum and microwave billiard is lost at frequencies above $f_{\rm max}=c_0/2h$ and, according to Weyl's law~\refeq{Weyl} the number of resonances is proportional to the area of the resonator and increases quadratically with frequency below $f_{\rm max}$. Thus, we needed to find a compromise between a large number of eigenfrequencies and a high $Q$ factor. Accordingly, we designed a sector microwave billiard with radius $R=800$~mm and height $h=20$~mm, corresponding to a cutoff frequency $f_{\rm max}=7.5$~GHz. Thereby, we achieved quality factors of several $1000$th and thus were able to determine, e.g., for the microwave billiard containing three scatterers $\approx 550$ eigenvalues. This was possible, because it exhibits GOE statistics implying that resonance spacings close to zero are most unlikely [see~\refeq{Wigner}], whereas for the empty microwave billiard the eigenfrequencies can be very close to each other in comparison to the average resonance spacing as the associated nearest-neighbor spacing distribution~\refeq{Poisson} is maximal for spacing zero, so that we could identify only 220 ones. Figure~\ref{fig:spec} shows a part of a transmission spectrum which contains isolated and weakly overlapping resonances.   

\begin{figure*}[htbp]
\begin{center}
\includegraphics[width=8.0cm,height=6.2cm]{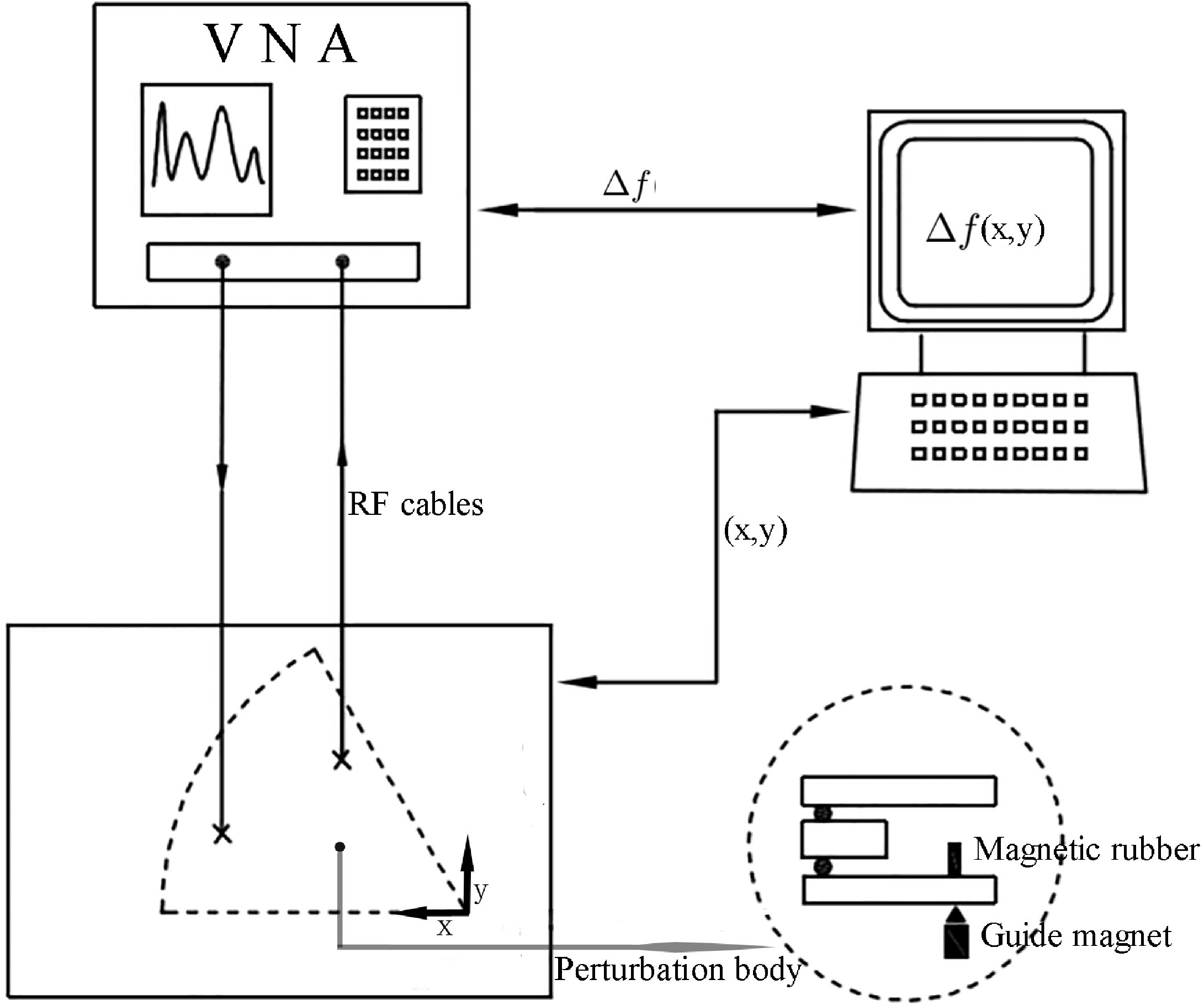}
\caption{
Schematic view of the experimental setup which was used for the measurement of the electric field intensity distributions at the eigenfrequencies of isolated resonances. A cylindrically shaped perturbation body made of magnetic rubber was inserted into the resonator thereby causing a frequency shift $\Delta f(x,y)$, which according to the Slater theorem is proportional to the squared electric field at its position. The electric field intensity distribution $E_{\rm z}^2(x,y)$ was determined by moving the perturbation body along the bottom plate of the resonator with an external guiding magnet, which was fixed to a positioning unit, and measuring $\Delta f(x,y)$.}
\label{fig:position}
\end{center}
\end{figure*}

The electric field intensity distribution was measured with the perturbation body method~\cite{Doerr1998,Dembowski1999} which is based on Slater's theorem~\cite{Maier1952a} stating that the frequency shift caused by introducing a metallic perturbation body into a microwave resonator leads to a frequency shift which depends on the difference of the squared electric and magnetic field,
\be
\Delta f(x,y)=f(x,y)-f_0=f_0\left(c_1 E_{\rm z}^2(x,y)-c_2\vec B^2(x,y)\right).
\ee
Here, $c_1$ and $c_2$ depend on the geometry and material of the perturbation body and $f_0$ denotes the resonance frequency of the resonator before introducing it. We are interested in the distribution of $E_{\rm z}(x,y)$ and, therefore, removed the contribution of $\vec B(x,y)$ by choosing a cylindrical perturbation body which was made from magnetic rubber (NdFeB)~\cite{Bogomolny2006}. The electric field intensity distribution was measured by moving the perturbation body along the resonator plane with an external magnet which was fixed to a positioning unit, as illustrated in Fig.~\ref{fig:position} Note, that with this method only the modulus of $E_{\rm z}(x,y)$ is accessible. To determine the phase, the perturbation body would need to be replaced by an antenna~\cite{Kuhl2007}. 

\section{Spectral fluctuation properties and wavefunctions\label{EV}}
The spectra were measured for frequencies $f\leq f_{\rm max}\simeq 7.5$~GHz in steps of 100 kHz. As mentioned above, the eigenfrequencies were determined by fitting the complex Breit-Wigner form~\refeq{Breit} to the measured scattering matrix elements $S_{\rm ba},\ a,\ b\in \{1,2\}$. For this to be feasible their precise experimental determination is indispensable, that is, all systematic errors need to be removed. The dominant contributions come from the coaxial cables connecting the VNA with the cavity, which attenuate and reflect the rf signal. These effects were removed by a proper calibration of the VNA before each measurement~\cite{Dembowski2005}. The experiments were performed at room temperature so that we had to deal with absorption, that is, weakly overlapping resonances. The fitting procedure might fail in cases, where the overlap is too strong or where two eigenfrequencies are lying too close to each other. To reduce these effects we optimized the quality factor of the cavity as outlined above. Still, since according to~\refeq{Weyl} the average resonance spacing decreases $\propto 1/f$ we could resolve the resonances only for frequencies below a frequency which was smaller than $f_{\rm max}$. Another cause for missing resonances are situations where the electric field strength is zero at the position of an antenna so that they cannot be excited. To avoid this, we performed the measurements for various positions of the antennas. In order to locate missing eigenfrequencies we looked at the fluctuating part of the integrated spectral density $N^{\rm fluc}(f_n)$, that is, the difference of the number of identified eigenfrequencies below $f_n$ and the expected number~\refeq{Weyl}, $N^{\rm Weyl}(f_n)$. At a missing eigenfrequency its local average exhibits a jump by $\simeq -1$. Then, we carefully inspected all reflection and transmission spectra in the corresponding frequency regions to check, whether we oversaw a resonance because of the overlap with neighboring ones which would lead to a bump in a resonance curve. By this procedure and due to the above listed provisions we were able to identify all except less than 4\% of the eigenfrequencies in the integrable case and less than 2\% in the other two cases.

\begin{figure*}
\begin{center}
\includegraphics[width=7.6cm,height=5.1cm]{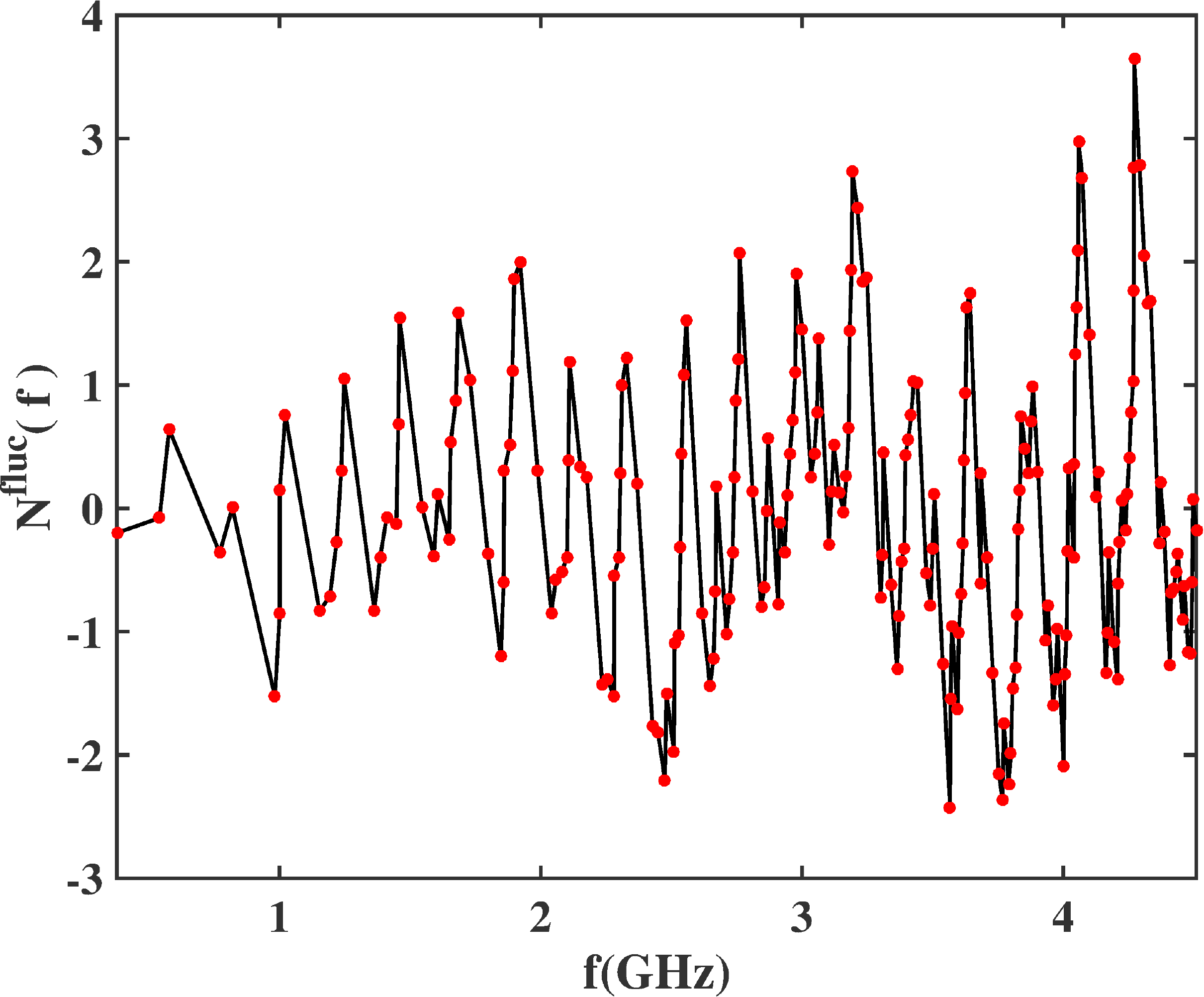}
\caption{ 
(color online) The fluctuating part of the integrated resonance density, $N^{\rm fluc}(f)=N(f)-N^{\rm Weyl}(f)$ in the frequency range 0.3-4.6 GHz (red dots). For better visibility they are connected by a black line.}
\label{fig:loss}
\end{center}
\end{figure*}
\begin{figure*}[h!]
\begin{center}
\includegraphics[width=8.cm,height=7.6cm]{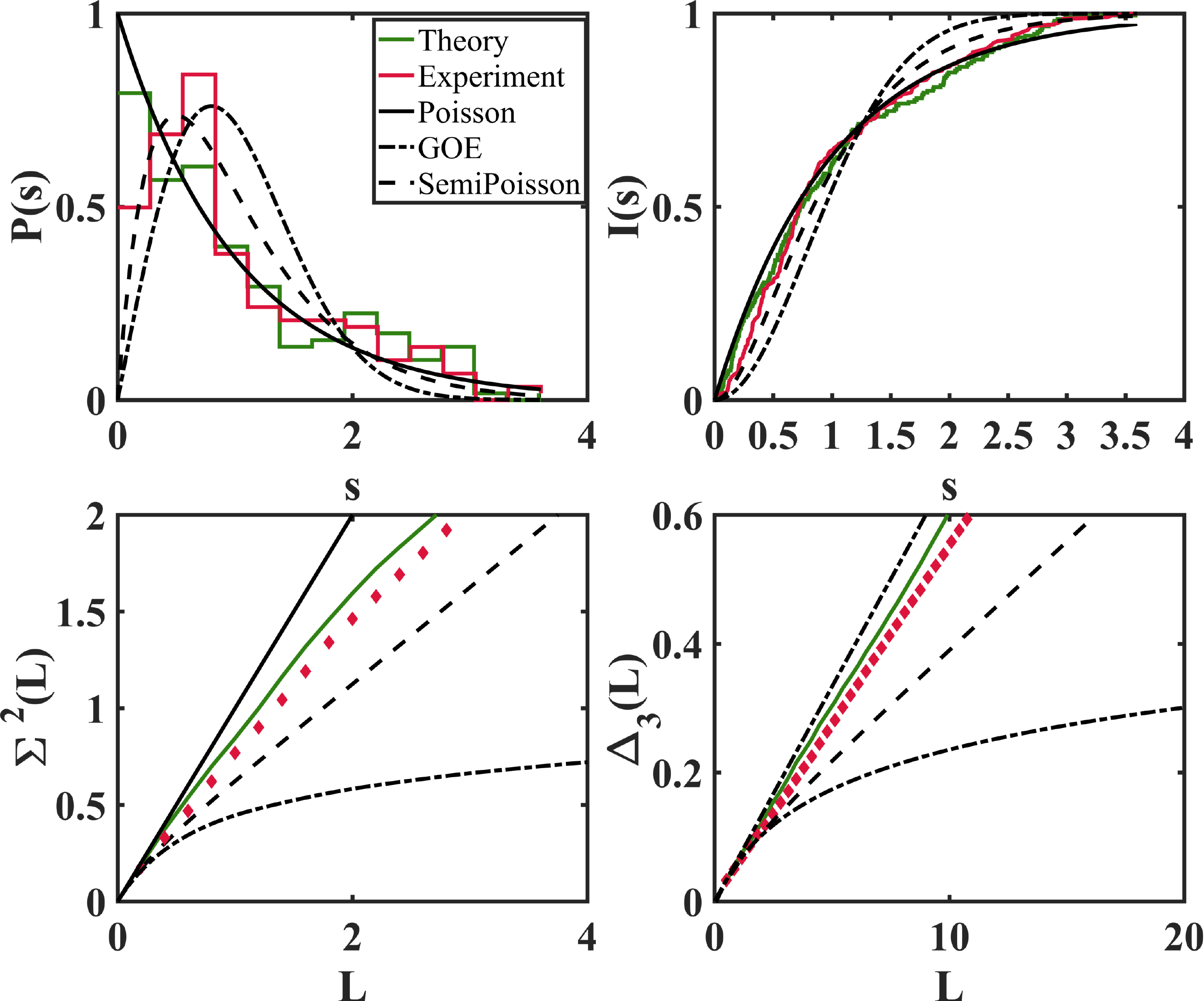}
\caption{
	(color online) Fluctuation properties of the unfolded eigenfrequencies of the empty microwave billiard. Shown are the nearest neighbor spacing distribution $P(s)$, the cumulative nearest-neighbor spacing distribution $I(s)$, the number variance $\Sigma^2(L)$ and the Dyson-Mehta statistics $\Delta_3(L)$. The solid, dashed and dash-dotted black lines show the curves for Poisson, GOE and semi-Poisson statistics, respectively. The red histograms and diamonds show the curves deduced from the measurements, the green histograms and solid lines were obtained from the computed eigenvalues; see~\refeq{eigenvalues}.}
\label{fig:empty}
\end{center}
\end{figure*}

One example for $N^{\rm fluc}(f_n)$, obtained from the eigenfrequency spectrum of an empty microwave billiard, is shown in Fig.~\ref{fig:loss}. Since the corresponding classical dynamics is integrable, there are many close lying eigenfrequencies. Therefore, we could resolve them only in the frequency range from 0.3-4.6~GHz and found about 220 eigenfrequencies, corresponding to 8 missing ones. For this case the eigenvalues and wavefunctions are known. We computed them using~\refeq{wavefunctions} and~\refeq{eigenvalues} and compared their spectral properties with those of the experimental data. The resulting curves, shown in Fig.~\ref{fig:empty} are close to each other. In order to reduce the deviations of the theoretical curves from Poisson behavior several 1000th of eigenvalues would be required. The observed, yet small, deviations between theory and experiment may have several causes. They cannot be attributed to missing levels since their effect on the spectral properties is negligible for quantum systems exhibiting Poisson statistics because the eigenvalues are either uncorrelated or only very weakly correlated. However, the Helmholtz equation of the empty microwave billiard including the antennas, which reached 8~mm into the cavity and thus were long and corresponded to two-dimensional dipoles with a frequency-dependent coupling to the resonator modes, is mathematically identical to the Schr\"odinger equation of a singular quantum billiard with the antennas corresponding to $\delta$-function potentials~\cite{Bialous2016,Tudorovskiy2011}. Their effect, nevertheless, is weak in the low-frequency range considered for the analysis of the spectral properties and thus causes only small deviations from the theoretical results as observed in Fig.~\ref{fig:empty}.   

In order to attain an almost-integrable system exhibiting intermediate statistics close to semi-Poisson statistics we inserted a copper disk, which had the same height as the resonator, into the microwave billiard. Rectangular quantum billiards containing finite-size circular scatterers were studied in detail in~\cite{Shigehara1996}. They demonstrated that the spectral properties may be similar to those of a pointlike scatterer, that is, of a quantum billiard containing a $\delta$-function potential, when the area of the scatterer is much smaller than that of the billiard and the wavelengths longer than the size of the scatterer so that they cannot resolve the shape of the scatterer. With increasing size of the scatterer wave chaos sets in at smaller and smaller energy values. This is to be expected, because it is well known that the classical dynamics of the Sinai billiard~\cite{Sinai1970}, or generally, of billiards with integrable shapes containing a finite-size scatterer~\cite{Rahav2003} are chaotic except for non-generic contributions resulting from waves which never hit the scatterer.

\begin{figure*}[h!]
\begin{center}
\includegraphics[width=8.0cm,height=7.6cm]{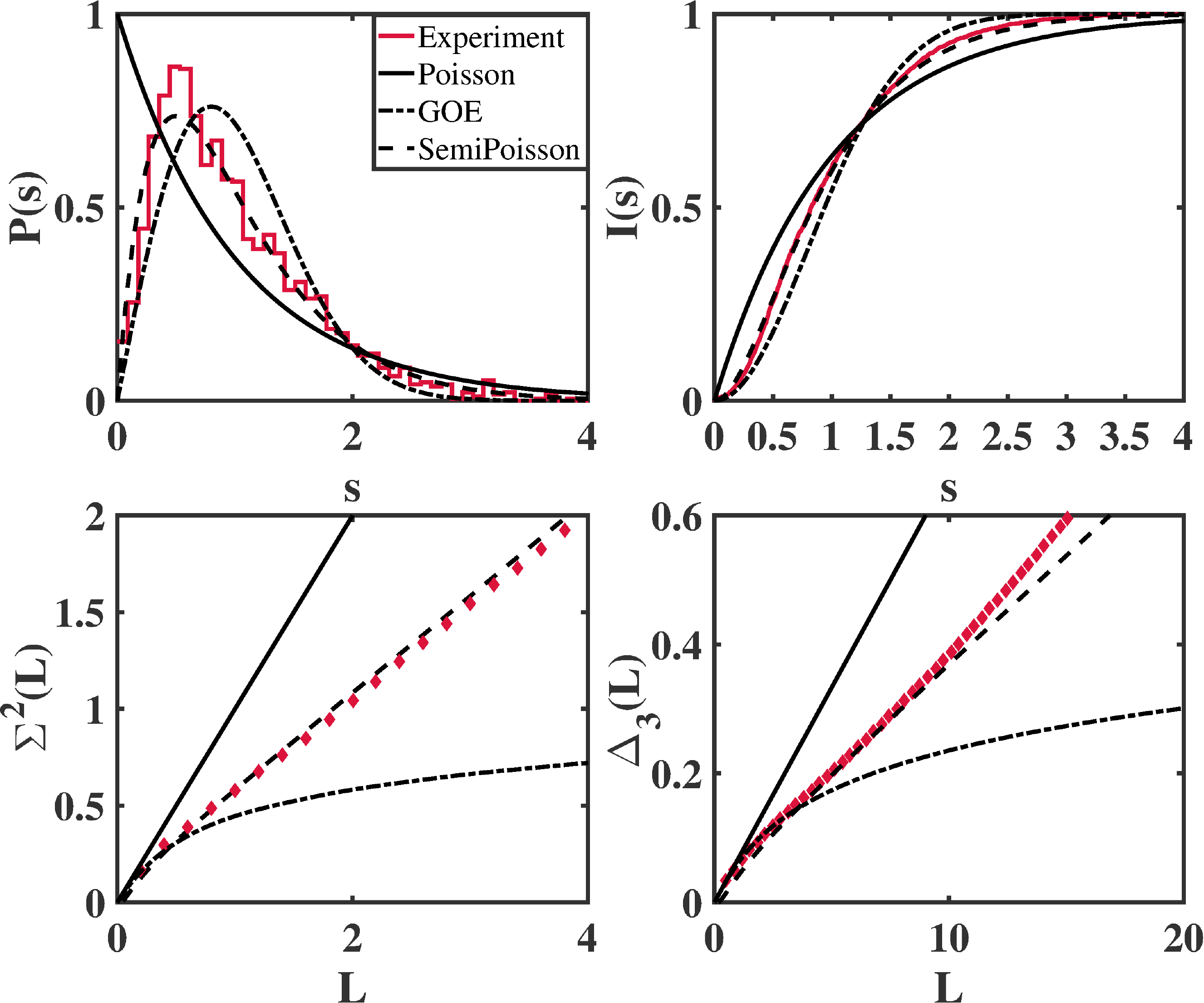}
\caption{
(color online) Fluctuation properties of the unfolded eigenfrequencies of the microwave billiard containing one disk. Shown is the average over the nearest neighbor spacing distribution $P(s)$, the cumulative nearest-neighbor spacing distribution $I(s)$, the number variance $\Sigma^2(L)$ and the Dyson-Mehta statistics $\Delta_3(L)$ for the different measurments listed in Tab.~\ref{tab:one}. The solid, dashed and dash-dotted black lines show the curves for Poisson, GOE and semi-Poisson statistics, respectively. The red histograms and diamonds show the curves deduced from the measurements.}
\label{fig:one}
\end{center}
\end{figure*}

\begin{figure*}[h!]
\begin{center}
\includegraphics[width=8.0cm,height=7.6cm]{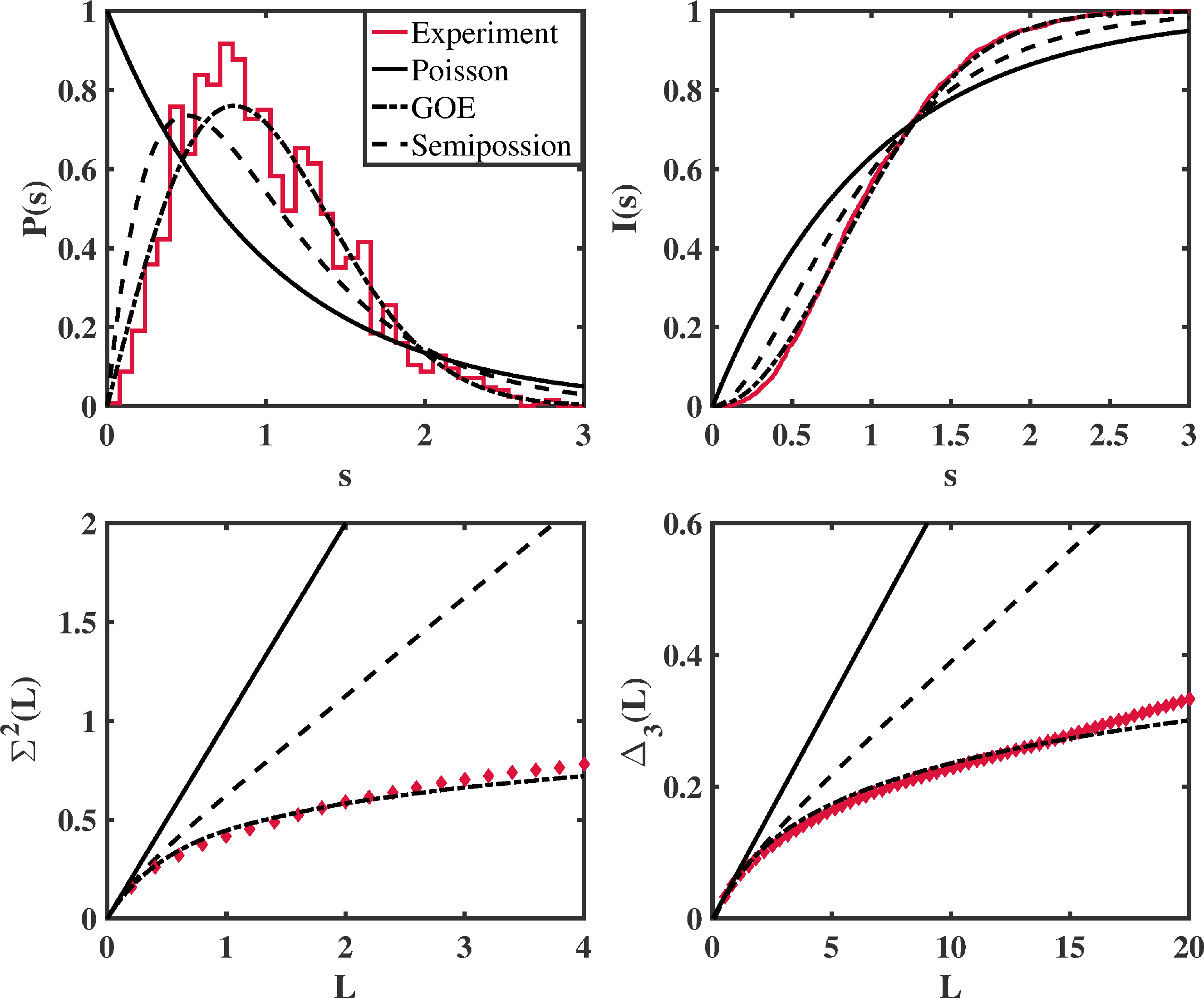}
\caption{
(color online) Same as Fig.~\ref{fig:one} for the experiments with three disks listed in Tab.~\ref{tab:three}.}
\label{fig:three}
\end{center}
\end{figure*}

The sizes and positions of the disks used in the experiments are listed in Tab.~\ref{tab:one}. All positions were chosen in the region of the curved boundary, because there most of the wavefunctions are non-vanishing, whereas in the region around the tip of the sector only $J_0$-type ones are non-vanishing below $f_{\rm max}$. Identifying eigenfrequencies was easier than in the integrable case, because the probability to find close-lying resonances was small, as clearly visible in the nearest-neighbor spacing distribution shown in Fig.~\ref{fig:one}. We determined them up to 6.94 GHz, however, because there were a few missing ones (less than 2\%), we split the eigenfrequency sequences into several complete ones before analysing the spectral properties. The total number of eigenfrequencies thus ranged between 300 and 500, which is sufficient to obtain statistically relevant results. In order to confirm that $P(s)$ indeed corresponds to semi-Poisson and not to a distribution intermediate between Poisson and GOE, we plotted it in a log-log plot, which clearly demonstrated that the decay indeed is exponential and not Gaussian-like. 
All statistical measures agree well with semi-Poisson statistics, as desired. Yet, we would like to emphasize, that this is not necessarily the case for almost-integrable or pseudo-integrable systems~\cite{Bogomolny2002,Shigehara1996}. Finally, in order to realize a quantum billiard exhibiting GOE statistics, we added three copper disks of the same height as the resonator. To obtain an ensemble of such systems we varied both size and position of the disks, as listed in Tab.~\ref{tab:three}. Eigenfrequency sequences were identified up to 7 GHz and then, as in the previous case, they were split into complete sequences comprising $\approx 300-550$ levels. The spectral properties indeed agree well with the corresponding GOE curves, as illustrated in Fig.~\ref{fig:three}.

\begin{figure*}[ht]
\begin{center}
\includegraphics[width=15.2cm,height=8.9cm]{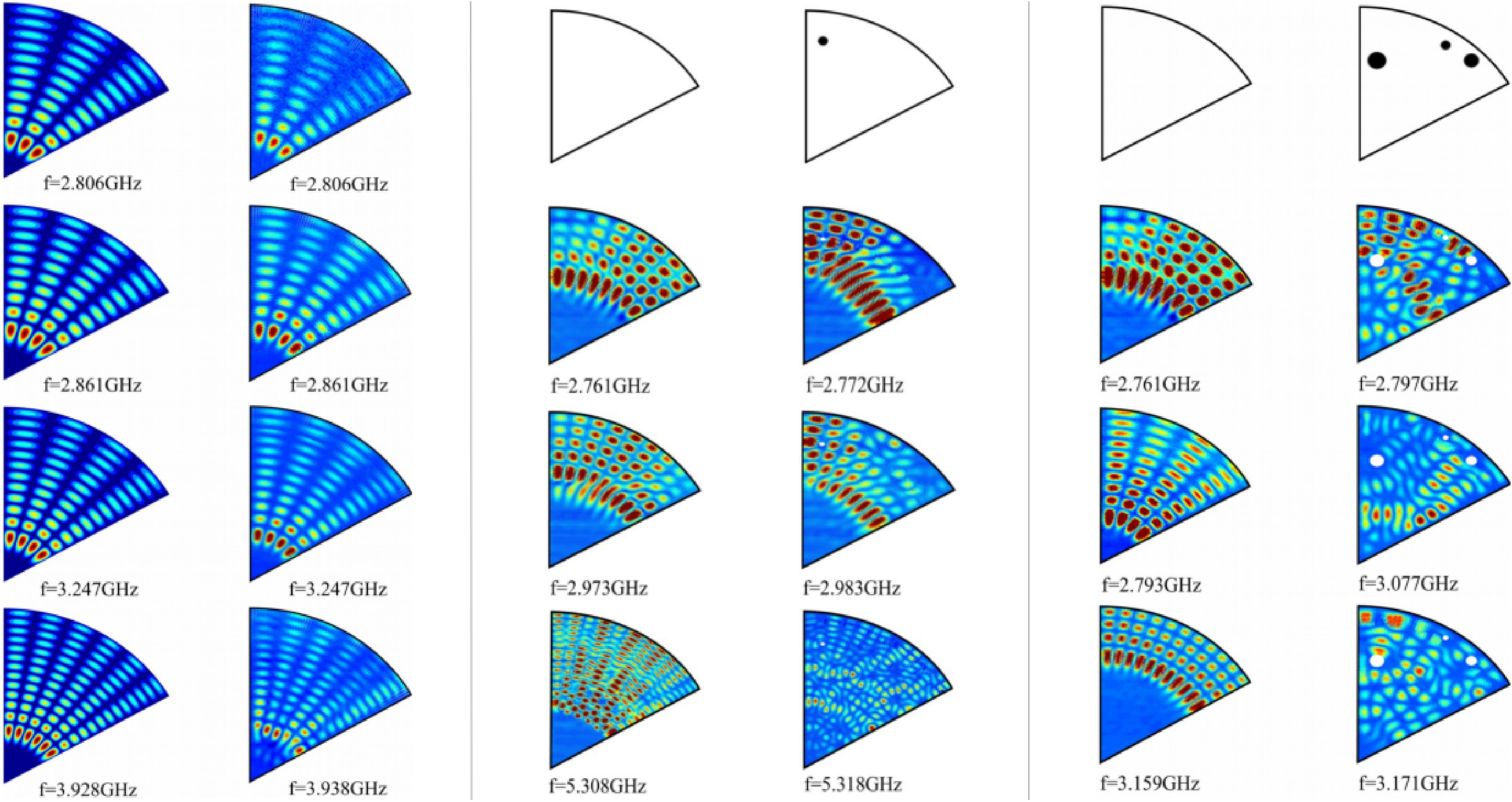}
\caption{
(color online) Measured electric field intensities. The first column shows computed wavefunctions, the second one the corresponding measured electric field intensities. The fourth and sixth columns show electric field distributions for a microwave billiard containing one (1st setup in Tab.~\ref{tab:one}) and three (5th setup in Tab.~\ref{tab:three}) scatterers, respectively. The corresponding distributions measured with no scatterer are shown in the third and fifth column.}
\label{WF}
\end{center}
\end{figure*}

We also measured electric field intensity distributions for a few eigenfrequencies corresponding to wavefunctions solving~\refeq{Schr}, where we tuned the frequency to that of well isolated resonances and then used the perturbation body method as outlined in Sec.~2. The first column in Fig.~\ref{WF}  shows computed wavefunctions for the $60^\circ$-sector billiard and the second one the corresponding electric field distributions. They agree very well, and thus demonstrate the precision of the wavefunction measurements and corroborate our assumption that the microwave billiard can be considered as a closed system even though it is a scattering system. In order to illustrate the effect of finite-size scatterers on the wavefunctions we compare electric field distributions of the empty microwave cavity shown in the third column to those obtained after inserting one copper disk of small size. The size of the scatterer, marked by a white circle in the latter, was much smaller than the billiard area. Nevertheless, it already shows a clearly visible effect at low frequencies leading to increasing distortion of the wavefunctions with increasing eigenfrequency, that is, decreasing wavelength. When adding three scatterers, the electric field intensity pattern of the empty microwave billiard, shown in the fifth column, is changed  considerably, as illustrated for the corresponding distributions in the sixth column. 

Because the wavefunction measurements are extremely time-consuming we only measured four of them for each of the three cases, i.e., for the microwave billiards containing no, one and three copper disks, respectively. However, for the statistical analysis of the distribution of the wavefunction components and their correlations, larger data sets are required. Nevertheless, we may obtain information on the properties of the wavefunctions from the widths and the amplitudes of the scattering matrix, which can be determined from the fit of the Breit-Wigner form~\refeq{Breit} to the expermental spectra. Indeed, the partial widths associated with the emitting and receiving antennas, $\Gamma_{\rm n a}$ and $\Gamma_{\rm n b}$ are proportional to the electric field intensity at their positions. They enter~\refeq{Breit} via the amplitudes $\sqrt{\Gamma_{\rm n a} \Gamma_{\rm n b}}$ and the resonance width $\Gamma_{\rm n}$, yet may not be determined individually~~\cite{Dembowski2005} because of the nonnegligible contribution of absorption to $\Gamma_{\rm n}$. As in this reference we therefore studied instead the distribution of the strengths $y_{\rm ba}=\Gamma_{\rm n b} \Gamma_{\rm n a}$ to gain insight into the statistical properties of the wavefunction components. For fully chaotic systems they are Gaussian distributed, and, accordingly, the partial widths have a Porter-Thomas distribution~\cite{Porter1965,Guhr1998,Dembowski2005}, implying that their product has a $K_0$ distribution, where $K_0$ is the zero-Bessel function of imaginary argument,
\be
P(y)=\frac{K_0\left(\sqrt{\frac{y}{\tau_{\rm a}\tau_{\rm b}}}\right)}{\pi\sqrt{\frac{y}{\tau_{\rm a}\tau_{\rm b}}}}\frac{1}{\tau_{\rm a}\tau_{\rm b}},
\ee
with $\tau_{\rm a}\tau_{\rm b}$ denoting the expectation value of $y$. Since $P(y)$ diverges for $y\to 0$ we transformed $y$ to $z=\log_{10}\left(\frac{y}{\tau_{\rm a}\tau_{\rm b}}\right)$. 
\begin{figure*}
\begin{center}
\includegraphics[width=7.2cm,height=5.1cm]{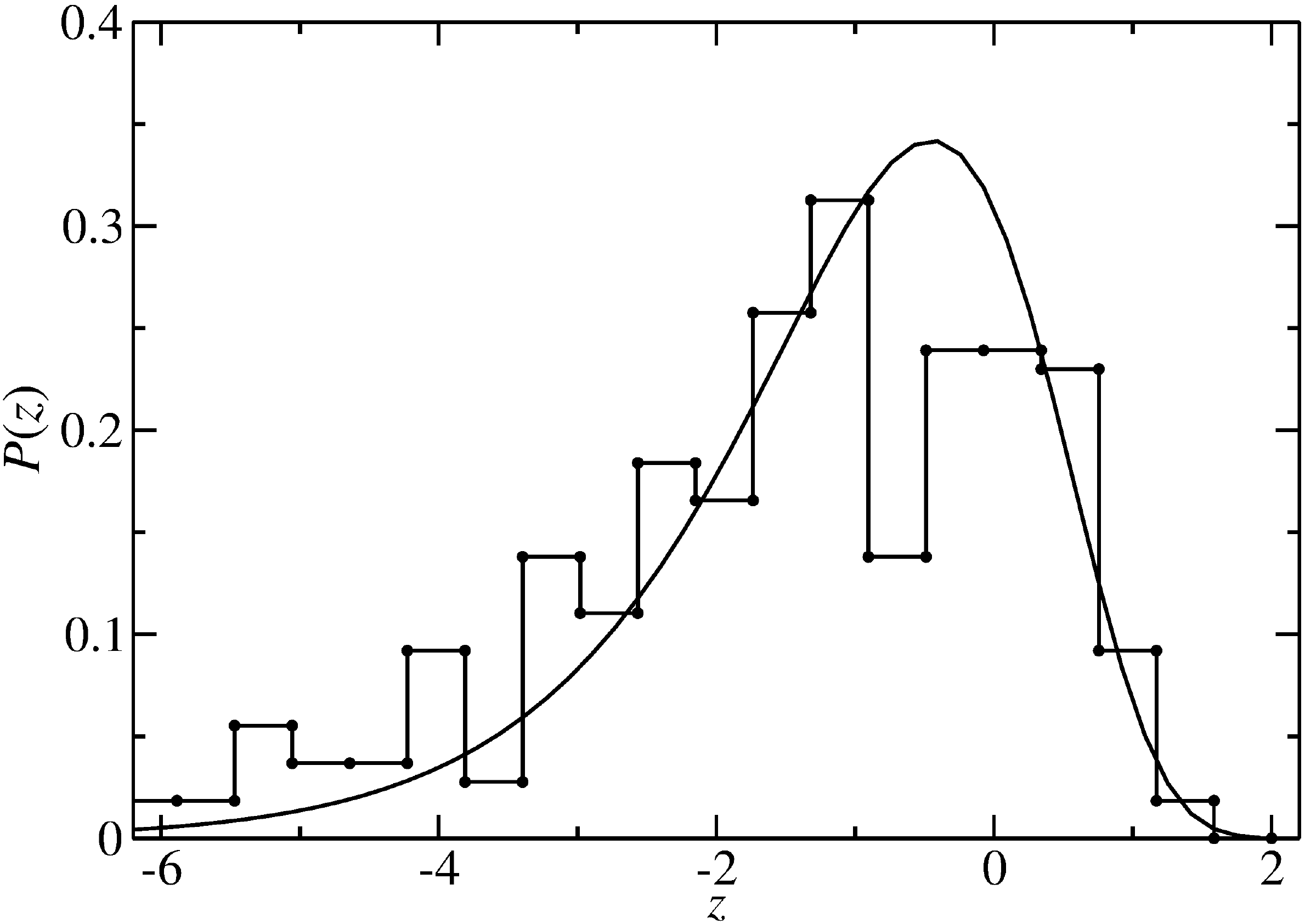}
\includegraphics[width=7.2cm,height=5.1cm]{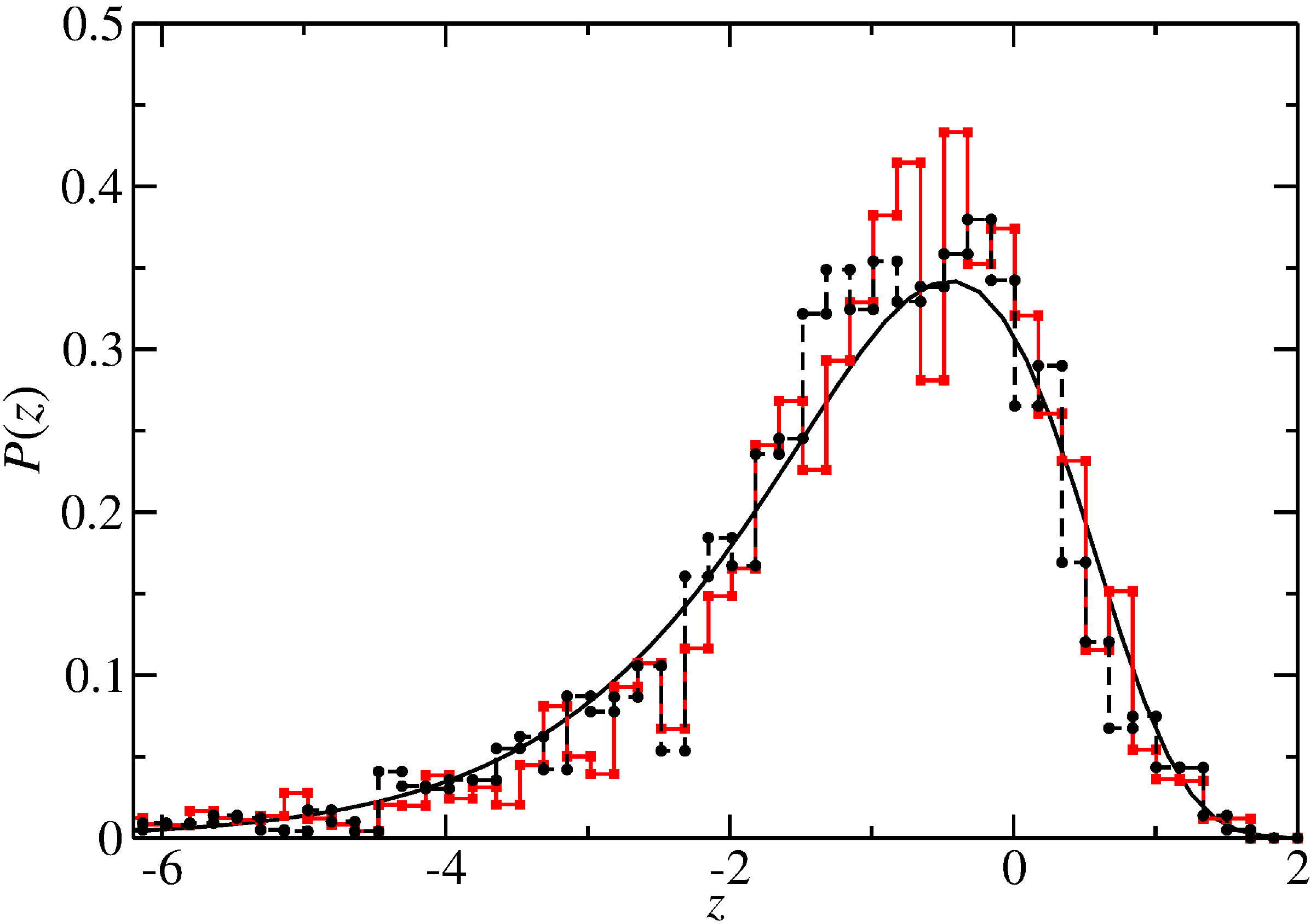}
\caption{ 
(color online) Experimental strength distribution (histogram) for the empty cavity (top) and for one (black dots) and three (red squares) added scatterers (bottom) in comparison to the GOE result (solid black line).} 
\label{K0Distr}
\end{center}
\end{figure*}
We determined the experimental strengths distributions by proceeding as in Ref.~\cite{Dembowski2005}. As expected~\cite{Dittes2000}, the results for the empty microwave cavity (histogram) shown in the top panel of Fig.~\ref{K0Distr} clearly deviate from a $K_0$ distribution. However, both for the almost-integrable (black) and the fully chaotic (red) cases shown in the bottom panel we find a good agreement with the $K_0$ distribution. This implies, that while the spectral properties of almost-integrable systems depend sensitively on the features of the classical dynamics, their strength distribution is close to that expected for typical, fully chaotic systems. Yet, in Ref.~\cite{Dembowski2005} clear deviations from the $K_0$ distribution were found for systems with a mixed regular-chaotic dynamics. This implies, that the insertion of a scatterer of which the size is much smaller than the area of the billiard, into the quantum or microwave billiard already induces in the low-frequency range, that is the long-wavelength region, strong distortions in the wavefunctions, which indeed are visible in Fig.~\ref{WF}, leading to features typical for chaotic systems. Note that while the distributions of the wave functions of almost-integrable systems were predicted to coincide with those of fully chaotic systems, this must not be the case for their spatial correlations~\cite{Seba1990,Seba1991}.

\section{Fluctuation properties of the scattering matrix\label{Spectra}}
Another focus of interest were the fluctuation properties of the scattering matrix of open systems with an integrable or almost-integrable dynamics in the scattering zone. Actually, since the resonance spectra are measured by emitting microwave power into the resonator via one antenna, thereby exciting an electric field mode in its interior, and receiving it at the same or another one, microwave billiards can also be viewed as scattering systems. Here, the antennas act as single-mode channels which couple the resonator modes to the exterior, and the resonator corresponds to the scattering zone, respectively. The scattering matrix formalism describing microwave resonators was shown to be identical with that for compound-nucleus reactions~\cite{Albeverio1996,Mahaux1969}. This analogy has been employed in the previous group of one of the authors (BD) in a sequence of experiments~\cite{Dietz2008,Dietz2009,Dietz2010,Dietz2010a,Kumar2013,Dietz2017,Kumar2017} to investigate universal properties of the scattering matrix for compound-nucleus reactions and, generally, for quantum scattering processes with intrinsic chaotic dynamics, that is, to verify analytical results derived on the basis of the supersymmetry and RMT approach.   

The scattering matrix approach~\cite{Mahaux1969} used for the derivation of RMT-based analytical expressions characterizing the fluctuation properties in the reflection and transmission spectra of a chaotic scattering system was developed by Mahaux and Weidenm{\"u}ller in the context of compound-nucleus reactions. The associated scattering matrix is given by
\begin{equation}
        S_{\rm ba}(f) = \delta_{\rm ba} - 2\pi i\left[\hat W^\dagger\left(f\II-\hat H^{eff}\right)^{-1}\hat W\right]_{\rm ba}.
\label{eqn:Sab}
\end{equation}
Here, $a$ and $b$ refer to the antenna channels and $\hat H^{eff}=\hat H-i\pi\hat W\hat W^\dagger$ with $\hat H$ simulating the spectral fluctuation properties of the Hamiltonian of the closed resonator or quantum billiard and $\hat W$ accounting for the coupling of the resonator modes to their environment. If the shape of the resonator coincides with that of a classically fully chaotic billiard, $\hat H$ is replaced by a random $N\times N$-dimensional matrix from the GOE. The matrix elements $W_{\rm a \mu}$ and $W_{\rm b \mu}$ describe the couplings of the antenna modes to the resonator modes. Furthermore, absorption in the wall of the resonator is modeled~\cite{Dietz2009,Dietz2010} by $\Lambda$ fictitious channels $W_{\rm c \mu}$. In the microwave experiments the frequency-averaged $S$-matrix was diagonal, $\langle S_{\rm ba}\rangle=\langle S_{\rm aa}\rangle\delta_{\rm ba}$, that is, direct processes were negligible. This property is accounted for in the RMT model through the orthogonality property $\sum_{\mu = 1}^N W_{\rm c \mu} W_{\rm c^\prime \mu}=N v_{\rm c}^2 \delta_{\rm cc^\prime}$. For $c=a,\, b$ the parameter $v^2_{\rm c}$ corresponds to the average strength of the coupling of the resonances to channel $c$, that is, the average size of the electric field at the position of the antenna. The input parameters of the RMT model~\refeq{eqn:Sab} are the transmission coefficients 
\be
T_{\rm c} = 1 - |\left\langle{S_{\rm cc}}\right\rangle |^2,
\label{Transm}
\ee
which provide a measure for the unitarity deficit of the average scattering matrix $\langle S\rangle$. They are related to $v^2_{\rm c}$ via $T_{\rm c} = \frac{4 \pi^2 v^2_{\rm c} / d}{(1 + \pi^2 v^2_{\rm c} / d)^2}$ with $d=\sqrt{\frac{2}{N}\langle H_{\rm\mu\mu}^2\rangle}\frac{\pi}{N}$ denoting the mean resonance spacing. 

\begin{figure*}
\begin{center}
\includegraphics[width=7.2cm,height=5.1cm]{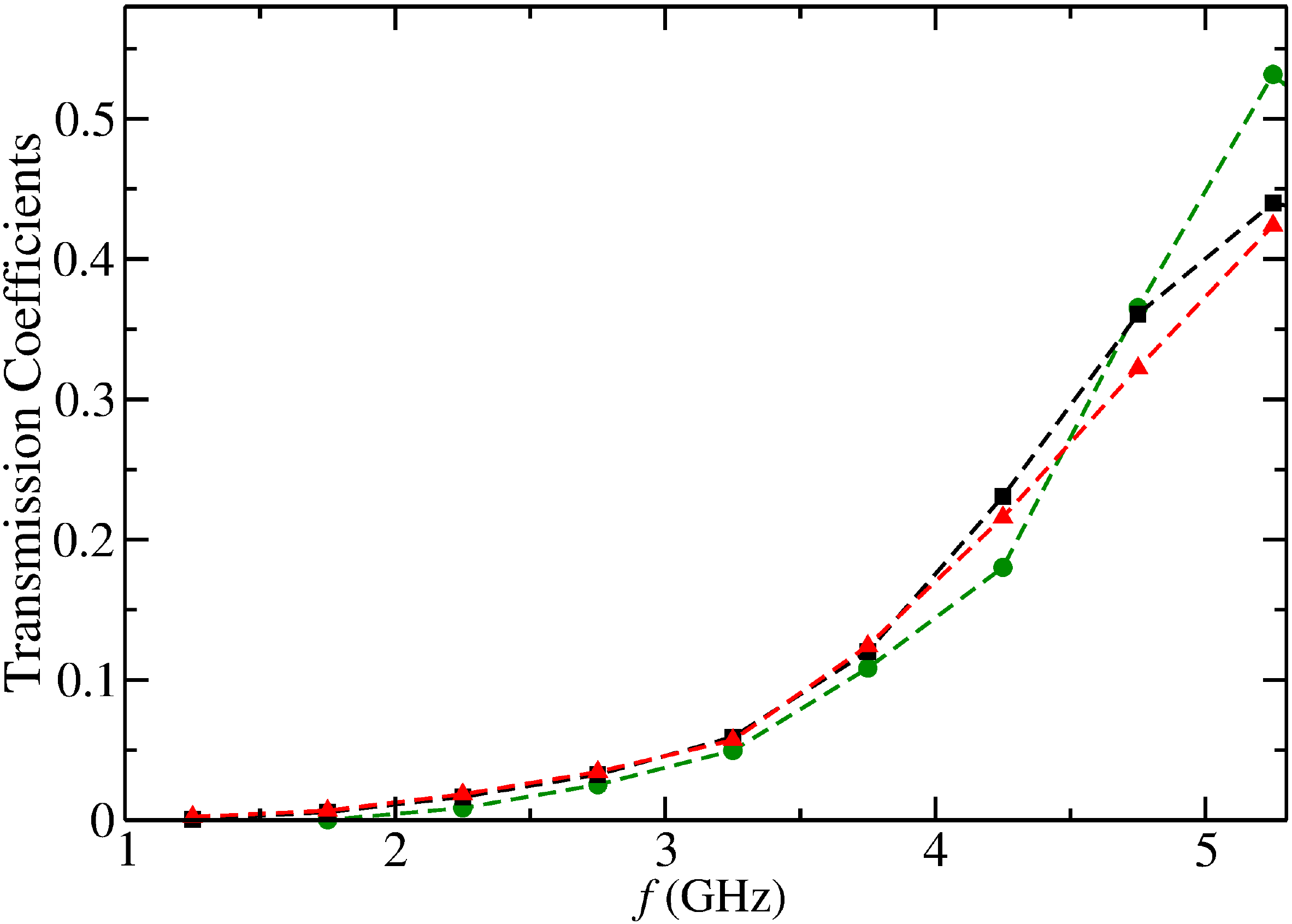}
\caption{
(color online) Variation of the transmission coefficients $T_{\rm c}$ associated with the antennas with frequency. They are nearly equal for both antennas and for varying number, positions and sizes of the scatterers. Therefore, the average over the respective ensembles composed of the realizations with, respectively, one and three disks, is shown for the average transmission coefficients for the cavities with no disk (green dots), one added disk (black squares) and three added disks (red triangles). To guide the eye of the reader the symbols are connected by dashed lines.}
\label{transcoeff}
\end{center}
\end{figure*}

The transmission coefficients $T_{\rm a}, T_{\rm b}$ associated with antennas $a$ and $b$ are obtained according to ~\refeq{Transm} from the measured reflection spectra, whereas those related to the fictitious channels, i.e., absorption, accounted for through the parameter $\tau_{\rm abs}=\Lambda T_{\rm c}$, need to be determined by fitting analytical results for the fluctuation properties like the two-point correlation function given in~\cite{Verbaarschot1985} or for the distribution of the scattering matrix elements~\cite{Dietz2010,Kumar2013} to the corresponding experimentally determined one~\cite{Dietz2010}. Because in the RMT model~\refeq{eqn:Sab} the coupling matrix $\hat W$ is assumed to be frequency independent we needed to ensure in the analysis of the experimental data that the resonance widths are approximately constant. Accordingly, we needed to divide the frequency range into windows of 0.5~GHz~\cite{Dietz2010}. 
An analytical expression was derived for the two-point correlation function of the scattering matrix elements,
\begin{equation}
        C_{ab}(\epsilon) = \langle S_{ab}(f)\,
        S_{ab}^\ast(f+\epsilon) \rangle - | \langle S_{ab}(f)
        \rangle |^2, \label{eqn:auto}
\end{equation}
in Ref.~\cite{Verbaarschot1985}. Furthermore, analytical expressions were derived for the distributions of the modulus $\vert S_{\rm ba}\vert$ and phase $\phi_{\rm ba}$ of the scattering matrix elements $S_{\rm ba}=\vert S_{\rm ba}\vert e^{i\phi_{\rm ba}}$ in Refs.~\cite{Fyodorov2005,Dietz2010,Kumar2013}. In order to determine the absorption parameter $\tau_{\rm abs}$ and to verify the values of the transission coefficients associated with the antennas computed with~\refeq{Transm}, we compared these analytical expressions to the corresponding experimental curves obtained for the microwave billiard containing three disks since its spectral properties follow GOE statistics. Figure~\ref{transcoeff} depicts the transmission coefficients associated with the antennas for the microwave billiards containing no (green dots), one (black squares) and three (red triangles) disks. Their values barely differ from each other below 5 GHz. Also the parameter $\tau_{\rm abs}$ should be similar in all three cases, as the absorption in the walls of the disks is negligibly small as compared to that in the walls of the cavity. Therefore, it makes sense to compare the RMT results obtained for the case with three disks with those for the cavities with no and one disk.  

\begin{figure*}[ht!]
\begin{center}
\includegraphics[width=7.2cm,height=7.2cm]{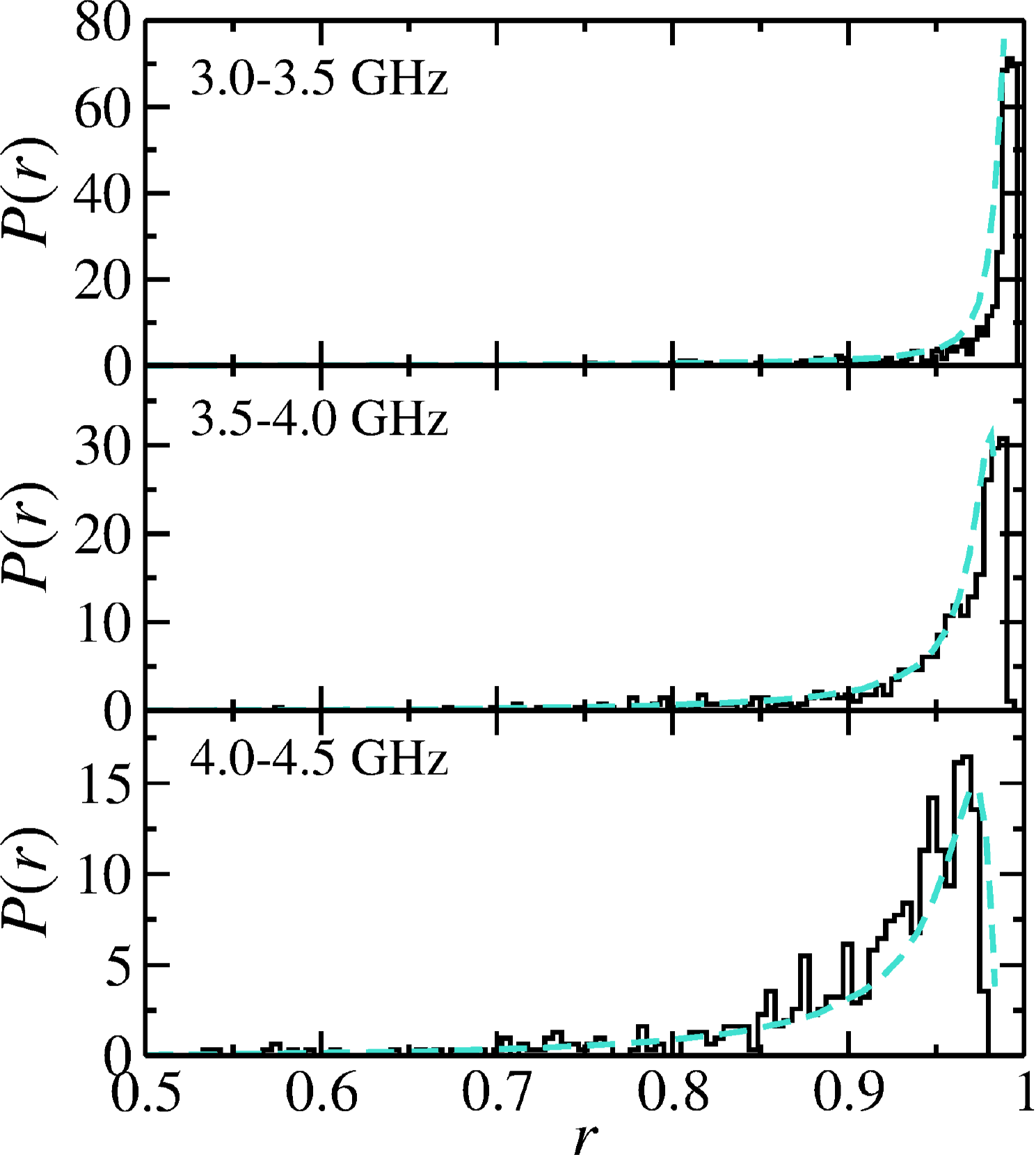}
\includegraphics[width=7.2cm,height=7.2cm]{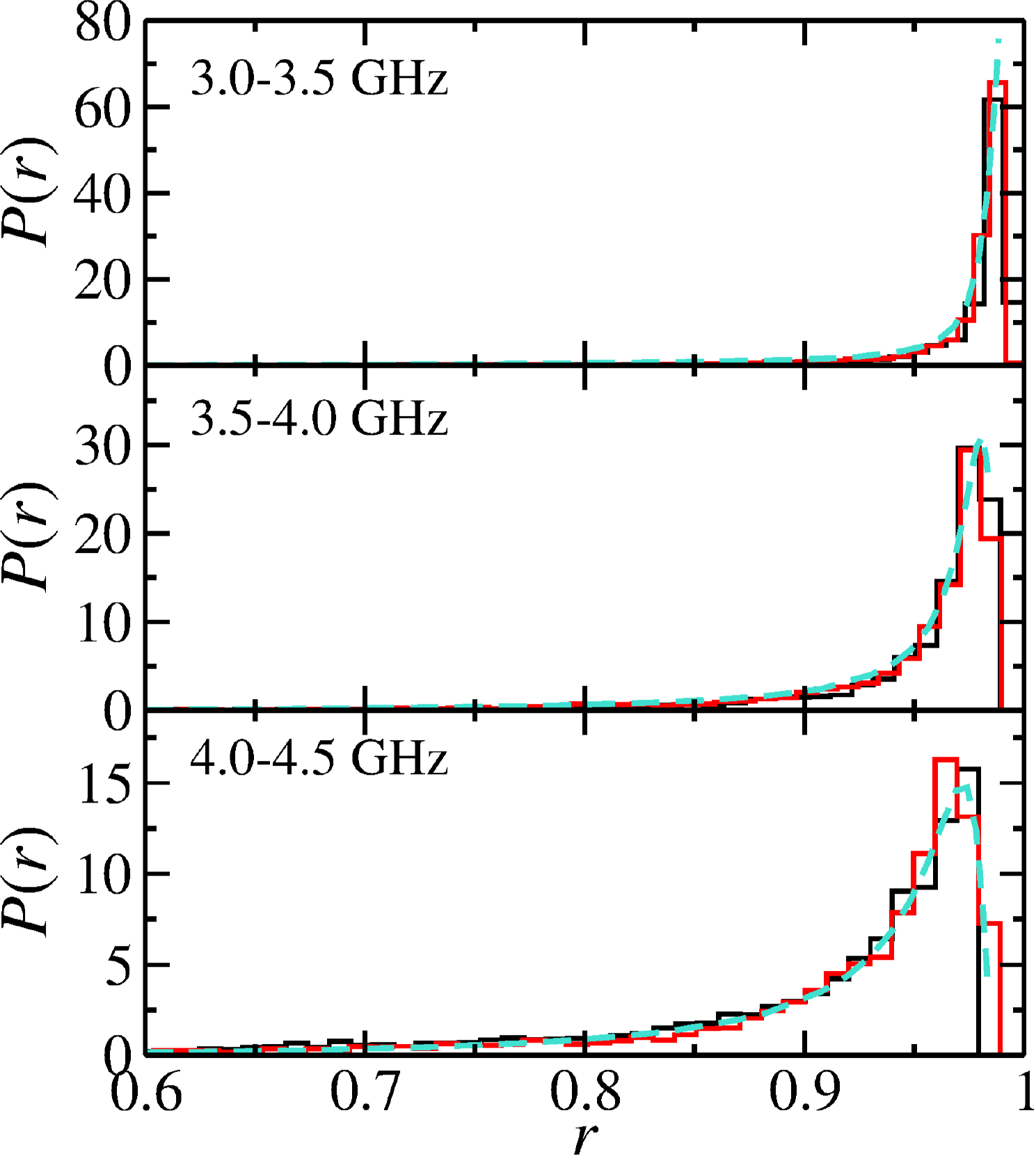}
\caption{
(color online) Experimental distribution of the modulus of the reflection matrix elements $r=\vert S_{\rm aa}\vert$ for the empty cavity (top,\ black histogram) and for one (black histogram) and three (red histogram) added scatterers (bottom) in comparison to the GOE result (turquoise dashed lines) obtained from a fit of the analytical distribution to that for the cavity with three disks. The transmission coefficients associated with the antennas were approximately the same and equaled $T_{\rm a}\simeq T_{\rm b}=0.091,\ 0.119,\ 0.165$ in the frequency ranges $[3.0,3.5],\ [3.5,4.0],\ [4.0,4.5]$~GHz, respecitvely.}
\label{ReflectDistr}
\end{center}
\end{figure*}

\begin{figure*}[ht!]
\begin{center}
\includegraphics[width=7.2cm,height=7.2cm]{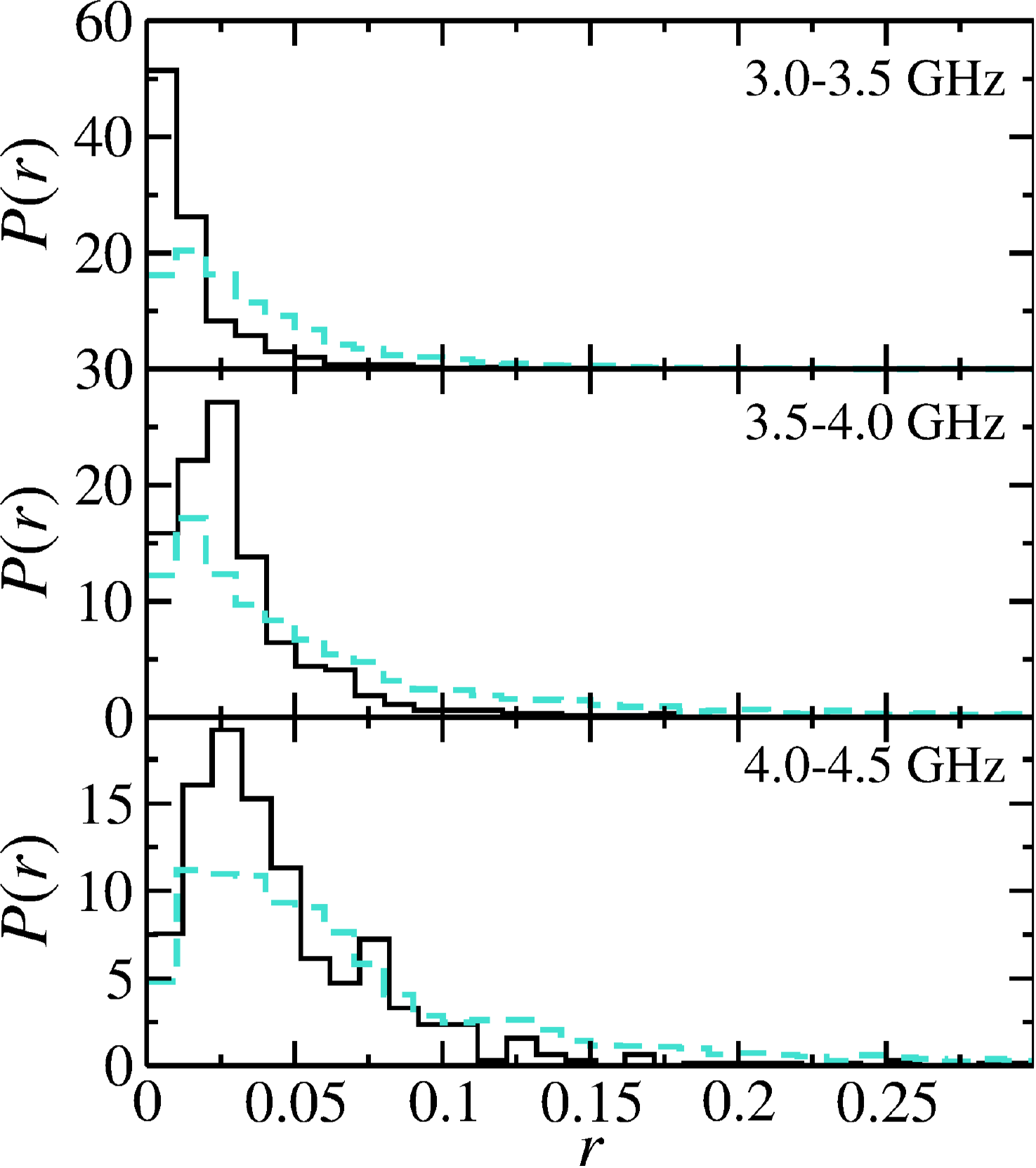}
\includegraphics[width=7.2cm,height=7.2cm]{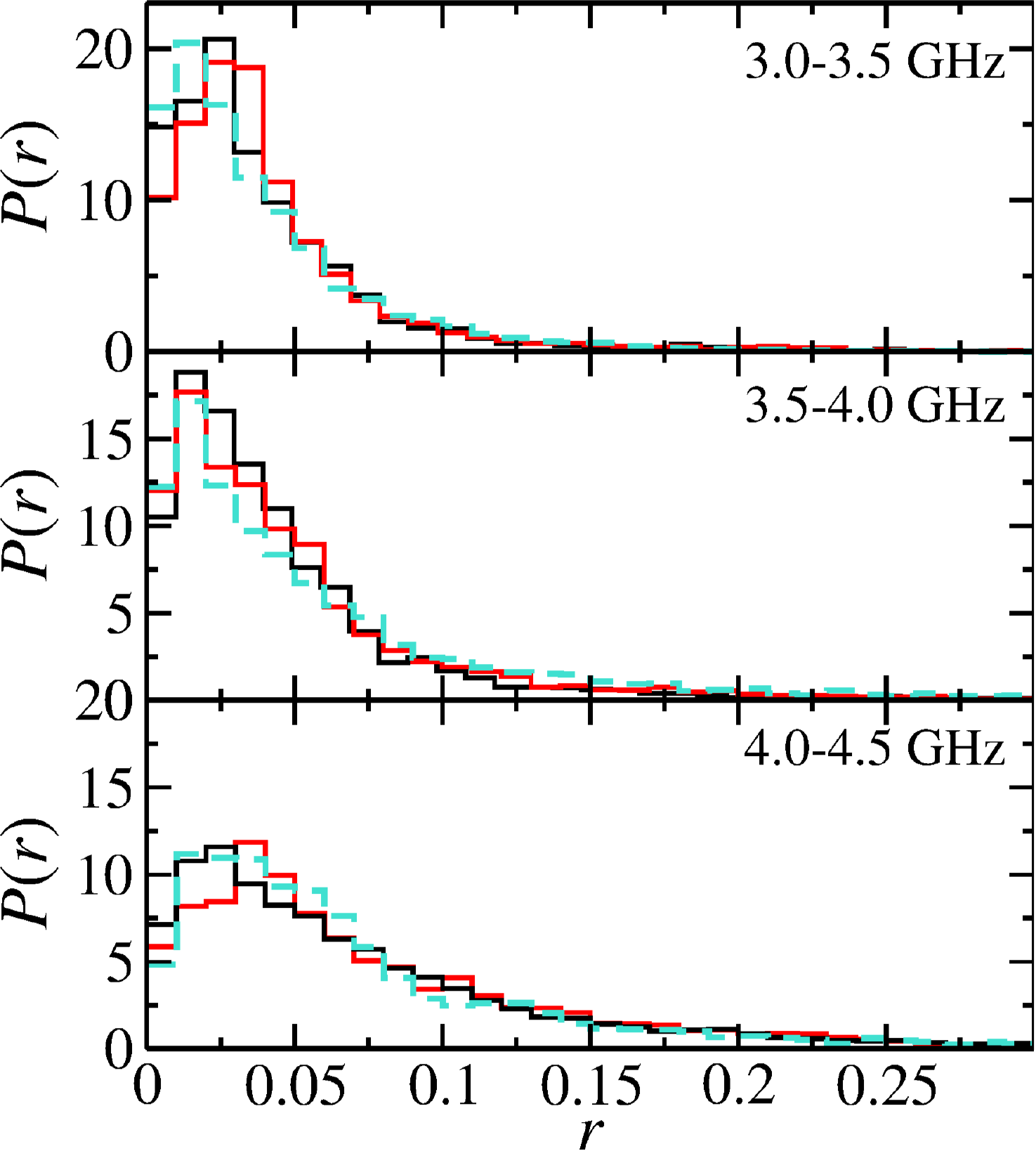}
\caption{
(color online) Same as Fig.~\ref{ReflectDistr} for the transmission matrix elements, $r=\vert S_{\rm ba}\vert ,\ a\ne b$}
\label{TransDistr}
\end{center}
\end{figure*}

\begin{figure*}[ht!]
\begin{center}
\includegraphics[width=7.2cm,height=7.2cm]{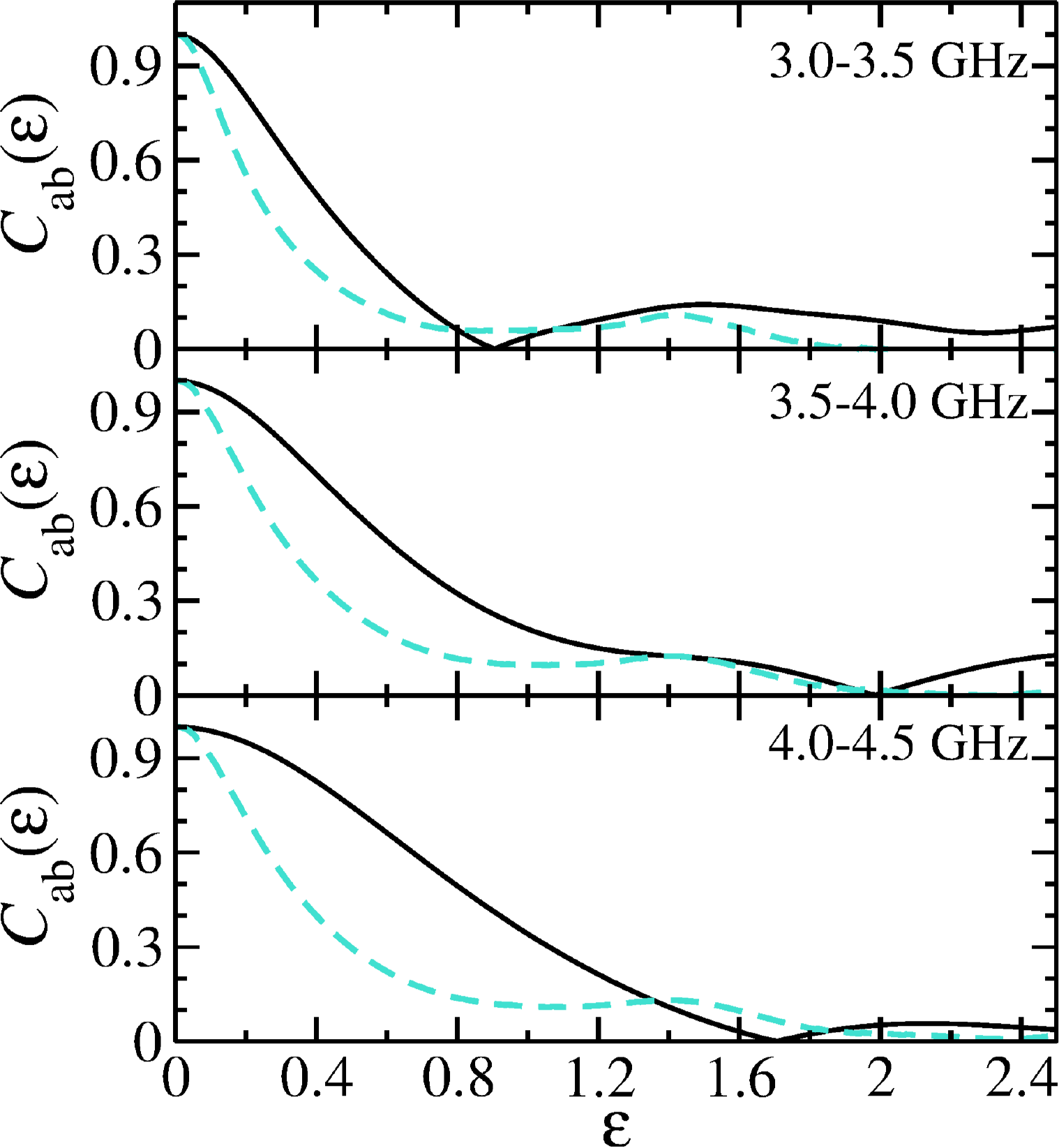}
\includegraphics[width=7.2cm,height=7.2cm]{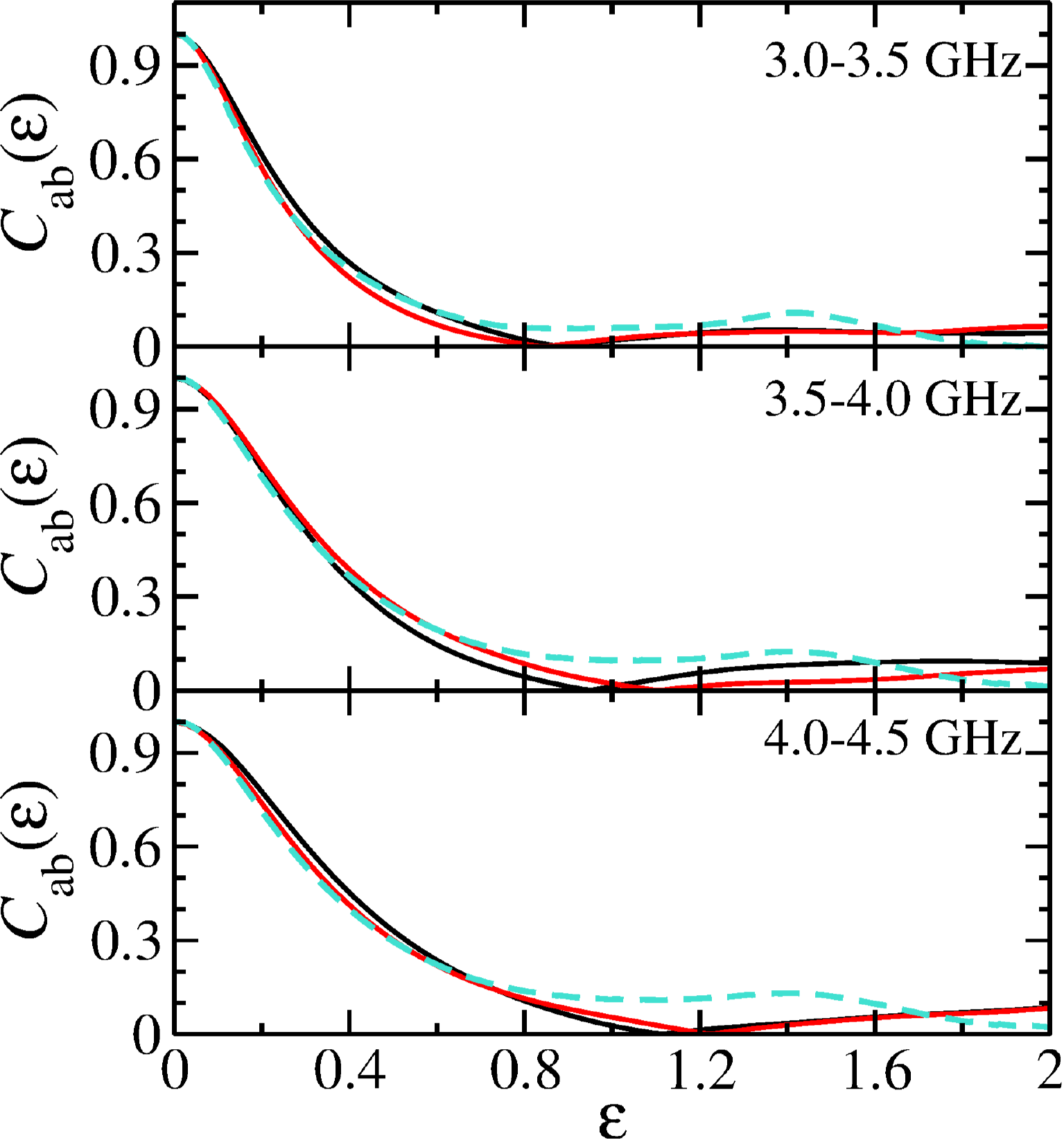}
\caption{
(color online) Experimental two-point $S$-matrix correlation functions for the empty cavity (top,\ black histogram) and for one (black histogram) and three (red histogram) added scatterers (bottom) in comparison to the GOE result (turquoise dashed lines).}
\label{Corr}
\end{center}
\end{figure*}

We, actually, determined the absorption parameter by fitting the analytical expression for the distribution of the modulus and phase of the reflection matrix elements $S_{\rm aa}$ to the experimental results and then inserted it into the other analytical expressions and RMT simulations based on~\refeq{eqn:Sab}. The resulting curves are shown as dashed turquoise lines in Fig.~\ref{ReflectDistr} together with the experimental distributions for the empty cavity for three different frequency ranges in the top panel, and for the cases with one (black histogram) and three disks (red histogram) in the bottom panel. The transmission coefficients associated with the antennas were approximately the same and equaled $T_{\rm a}\simeq T_{\rm b}=0.091,\ 0.119,\ 0.165$ in the frequency ranges $[3.0,3.5],\ [3.5,4.0],\ [4.0,4.5]$~GHz, respectively.     
Figure~\ref{TransDistr} exhibits the corresponding results for the transmission matrix elements $S_{\rm ba}$. Here, we performed RMT simulations using~\refeq{eqn:Sab}, since the analytical expressions for the distributions are even more complex for transmission than they are for reflection. In all cases the distributions for the empty cavity deviate considerably from the analytical ones, as expected~\cite{Dittes2000}, because the corresponding classical dynamics is integrable, whereas the RMT-based results are applicable to fully chaotic systems. Yet, both the curves for one disk and for three disks agree well with the analytical results, even though the classical dynamics of the former is almost-integrable and not fully chaotic. 
In Fig.~\ref{Corr} we show the autocorrelation function for the cases with no (top panel), and one (black) and three (red) disks (bottom panel) together with the analytical result which was obtained by inserting the values of $T_{\rm a},\ T_{\rm b}$ and $\tau_{\rm abs}$ deduced from the analysis of the distributions of the scattering matrix elements for the case with three disks (turquoise dashed line). Again, there is no agreement between the analytical and experimental results for the integrable case, whereas the curves lie on top of each other for the almost-integrable and chaotic ones. From these observations we may conclude that already a disk of a size which is small compared to the billiard area, the fluctuation properties of the scattering matrix are strongly affected, as is the strength distribution.

\section{Conclusions}
We experimentally investigated the properties of the eigenvalues, wavefunctions and of the scattering matrix of microwave billiards corresponding to billiards with integrable, almost-integrable and chaotic dynamics. For this we chose a billiard with the shape of a $60^\circ$ circle sector and added one and three disks, respectively. For the almost-integrable case we chose the size and position of the disk such that the spectral properties of the corresponding quantum system agreed well with semi-Posson statistics. While the spectral properties of pseudo-integrable and almost-integrable systems have been investigated extensively during the last two decades, the properties of the scattering matrix for an open system of which the dynamics in the scattering zone is pseudo-integrable~\cite{doCarmo2019} or almost integrable are not yet fully understood. Yet, microwave billiards provide an ideal system for such investigations since they correspond to scattering systems with the antennas acting as single-scattering channels and the classical dynamics in the scattering zone defined by its shape~\cite{Dietz2010}. Accordingly, we analyzed the fluctuations in the transmission and reflection spectra, that is, of the associated scattering matrix, which are known to be universal if the dynamics in the scattering zone is fully chaotic, and also the strength distribution which provides information on the statistical properties of the wavefunction components. For this we employed analytical results, which were obtained in the context of compound-nucleus reactions, and compared them first to the corresponding experimental results for the chaotic case in order to obtain the parameters characterizing the fluctuation properties of the scattering matrix, and then compared them to the integrable  and almost-integrable cases. Large deviations were observed in the former case, whereas good agreement was found for the latter one. This implies that the change from integrable to almost-integrable by introducing a singular scatterer turns the fluctuation properties of the scattering matrix from those typical for integrable systems to those for fully chaotic ones. These findings imply that the fluctuation properties of the scattering matrix and also of the strength distribution may not serve as a measure to distinguish between almost-integrable and chaotic classical dynamics based on purely quantum properties, whereas the spectral properties clearly discriminate between them.  
\section*{Acknowledgment}
We thank the National Natural Science Foundation of China for financial support under Grant Numbers 11775100 and 11775101. Furthermore, we would like to thank Florian Sch\"afer who helped us with the implementation of the fitting procedure and Maksym Miski-Oglu for his helpful advices during the setup of the experiments.

\end{document}